\documentclass[aps,twocolumn,showpacs,preprintnumbers,nofootinbib,prd,superscriptaddress,groupedaddress,10pt]{revtex4-1}

\makeatletter
\def\l@subsubsection#1#2{}
\def\l@subsubsubsection#1#2{}
\makeatother

\usepackage{graphicx,amssymb,amsmath,amsthm,amsfonts,epsfig,epsf,fixmath}
\usepackage[usenames]{color}
\usepackage{epstopdf}

\usepackage{bm}
\usepackage{amsmath}
\usepackage{dcolumn}
 \usepackage[latin1]{inputenc}
\usepackage{latexsym}
\usepackage{rotating}
\usepackage{longtable}

\usepackage{enumerate}
\usepackage{tensor,multirow}
\usepackage{mathtools}
\usepackage{url}
\usepackage[linktocpage]{hyperref}

\begin{document}
\newcommand{\IUCAA}{Inter-University Centre for Astronomy and
  Astrophysics, Post Bag 4, Ganeshkhind, Pune 411 007, India}
\newcommand{\INFN}{INFN Sezione di Ferrara, Via Saragat 1, 44122 Ferrara, Italy}  
\title{Effect of superfluid matter of a neutron star on the tidal deformability}
\author{Sayak Datta}\email{skdatta@iucaa.in} 
\affiliation{\IUCAA}
\author{Prasanta Char}\email{char@fe.infn.it}
\affiliation{\INFN}
\date{\today}
\begin{abstract}

 We study the effect of superfluidity on the tidal response of a neutron star in a general relativistic framework. In this work, we take a dual-layer approach where the superfluid matter is confined in the core of the star. Then, the superfluid core is encapsulated with an envelope of ordinary matter fluid which acts effectively as the low-density crustal region of the star. In the core, the matter content is described by a two-fluid model where only the neutrons are taken as superfluid and the other fluid consists of protons and electrons making it charge neutral. We calculate the values of various tidal love numbers of a neutron star and discuss how they are affected due to the presence of entrainment between the two fluids in the core. We also emphasize that more than one tidal parameter is necessary to probe superfluidity with the gravitational wave from the binary inspiral.

\end{abstract}

\maketitle

\section{Introduction}

The observation of a gravitational wave (GW) from the binary neutron star (BNS) merger event GW170817 has allowed us to study the physics of the extreme environment of highly dense matter at strong gravity  \cite{Abbott2017,Abbott2018}. During the orbital evolution, the tidal interaction between the stars of the binary deforms both of them. These deformations can be measured in terms of the relativistic tidal Love numbers of the stars \cite{Flanagan2008, Hinderer2008, Damour2009, Binnington2009, Hinderer2010}. Precise measurements of these parameters from the GW signal during the inspiral phase can be extremely useful to study the nature and the equation of state (EOS) of the supranuclear matter inside a neutron star (NS) \cite{Agathos2015, Takami2014, Bose2018}. This is why a huge effort has been made to understand the modification of waveforms due to the tidal Love numbers and their measurability and distinguishability of different EOSs \cite{Vines2011, Damour2012, Read2013, DelPozzo2013, Wade2014, Favata2014, Hotokezaka2016}. Moreover, one can also infer on the fluid nature of those objects. As these stars are supposedly very old, their core temperature should be below the critical transition temperature for the BCS-like pair formation \cite{Sedrakian2018}. Therefore, one can expect superfluid (SF) neutrons and superconducting protons to form at the core of the star and superfluid neutrons in the inner crust \cite{Migdal1959, Clark1992}. Pulsar glitches and the rapid cooling of the NS in Cassiopeia A are examples which are explicable invoking superfluid matter inside NS \cite{Baym1975, Anderson1975, Page2011, Shternin2011}. These changes in the fluid nature of the star from a single-fluid to a  multi-fluid object can influence its deformability in a non-trivial way \cite{Char2018}. Recently, we have investigated the role of superfluidity for the $\ell = 2 $ electric-type tidal Love number $k_2$ and the corresponding tidal deformability $\Lambda_{k_2}$ \cite{Char2018}, (hereafter, paper I). In this work we have modeled the star as a non-rotating sphere of superfluid nuclear matter. We had adopted the two-fluid model where one fluid is the neutron superfluid and the other is the normal charge-neutral fluid comprising protons and electrons \cite{Carter1989, Comer1994, Carter1995, carter1998_1, carter1998_2, Langois1998, Prix2000}. We found that the inclusion of superfluidity manifests significant change in $\Lambda_{k_2}$ compared to the non-superfluid case. 

However, a neutron star is also a multi-layered object i.e. the phases of matter differ significantly from the crust to the core. As has been known that the property of low density nuclear matter is correlated directly with the radius, one has to take into account a proper crust model in the calculation. To do so, we follow the method described in Ref.\cite{Andersson_2002}, where the properties of the superfluid region inside the core is appropriately matched to the normal fluid envelope encapsulating the core. Therefore, the superfluid neutrons are confined in the core where as the envelope acts as the low density region of the star. Although, we do not consider the elasticity of the crustal region in our formalism, this dual-layer core-envelop approach can approximate the structure of the star with a crust. Since crustal elasticity does not bring considerable change in the Love numbers it is unnecessary to include it here\cite{Biswas2019elasticity}. We also study the junction conditions for the perturbed quantities of interest in detail. 

At this point, it is important to note that when we speak of the deviation of $\Lambda_{k_2}$ due to the superfluid nature, we bring an ambiguity in our interpretation of the observed $\Lambda_{k_2}$. The value of $\Lambda_{k_2}$ in two-fluid calculation for a particular EOS model can be similar to the value in a single-fluid calculation for another EOS. So, we cannot distinguish between the EOS and also probe the fluid nature of matter at the same time with the measurement of $\Lambda_{k_2}$. One possible way to break the degeneracy is to have measurements of other Love numbers which have much smaller effects on the waveform. This gives us a primary motivation to study higher order electric-type Love numbers and magnetic-type Love numbers in the case of a superfluid star. 

The paper is organized as follows. In Sec. \ref{method}, we first
discuss the two-fluid formalism followed by the calculation
of the equilibrium structure along with a brief overview of the
RMF model of dense matter to calculate the assorted matter
coefficients of the model. Next, in Secs. \ref{even perturbation} and \ref{odd perturbation}, we derive the framework for even and odd parity tidal perturbations in the two-fluid model respectively. In Sec. \ref{Love} we discuss how the tidal Love numbers are calculated.  Then, in Sec. \ref{res} we discuss our results. We assume $c = G = 1$ and use the metric signature $(-,+,+,+)$ throughout the article.

\section{General relativistic superfluid neutron star}
 \label{method}

The main ingredients of the superfluid formalism have been developed and discussed in several works \cite{Carter1989,Comer1994,Carter1995,carter1998_1,carter1998_2,Langois1998,Comer1999,Prix2000,Andersson2001}. To incorporate SF matter inside NSs we follow a two-fluid model with entrainment. The central quantity of this formalism is the master function, $\Lambda$. It depends on three scalars, $n^2 = -n^{\mu}n_{\mu}$, $p^2 = -p^{\mu}p_{\mu}$, and $x^2 = -n^{\mu}p_{\mu}$, where $n^{\mu}$ and $p^{\mu}$ are the number density currents of the neutron and proton, respectively. When the fluids are co-moving,  $-\Lambda (n^2,p^2,x^2)$ represents the total thermodynamic energy density. The energy-momentum tensor takes the following form,
\begin{equation}
\label{SF EMT}
T^{\mu}_{\nu} = \Psi \delta^{\mu}_{\nu} + p^{\mu}\chi_{\nu} + n^{\mu}\mu_{\nu},
\end{equation}
where, $\Psi$ is the generalized pressure, and it can be expressed as,
\begin{equation}
\label{generalised pressure}
\Psi = \Lambda - n^{\rho}\mu_{\rho}-p^{\rho}\chi_{\rho},
\end{equation}
where, $\chi_{\nu}$ and $\mu_{\nu}$ are,  respectively, the chemical potential co-vectors of the proton and the neutron fluids.
\begin{equation}
\label{momentum covector eqn}
\mu_{\mu} = {\cal B} n_{\mu} + {\cal A} p_{\mu}, ~~~~ \chi_{\mu} = {\cal C} p_{\mu} + {\cal A} n_{\mu},
\end{equation}
where the ${\cal A, B}$ and ${\cal C}$ coefficients are defined as follows,
\begin{equation}
{\cal A} = -\frac{\partial\Lambda}{\partial x^2},~~~ {\cal B} = -2\frac{\partial\Lambda}{\partial n^2},~~~ {\cal C} = -2\frac{\partial\Lambda}{\partial p^2}.
\end{equation}
The expressions for $\mu_{\mu}$ and $\chi_{\mu}$ in Eq. \ref{momentum covector eqn} make the entrainment effect vivid. Momentum of the one component carries along some of the mass current of the other component when ${\cal A} \neq 0$. Thus, if ${\cal A} = 0$ the master function becomes ``entrainment-free" , implying that it is independent of $x^2$. 
The conservation equation for $n^{\mu}$ and $p^{\mu}$ implies,
\begin{equation}
\label{conservation of number}
\nabla_{\mu}n^{\mu} = \nabla_{\mu}p^{\mu} = 0.
\end{equation}
They also satisfy a set of Euler type equations \cite{Comer1999},
\begin{equation}
n^{\mu}\nabla{_{[\mu}\mu_{\nu]}} = p^{\mu}\nabla{_{[\mu}\chi_{\nu]}} = 0,
\end{equation}
where, the square brackets represent the antisymmetrization of the closed indices.

\subsection{Equation of state of nuclear matter}
\label{matter}

We have calculated the master function $(\Lambda)$ using the $\sigma$-$\omega$-$\rho$ model with self-interaction in the RMF approximation \cite{Comer2003, Comer2004, Kheto2014, Kheto2015}. The Lagrangian of the theory is as follows,

\begin{widetext}
\begin{eqnarray}\label{lag}
{\cal L}_B &=& \sum_{B=n,p} \bar\Psi_{B}\left(i\gamma_\mu{\partial^\mu} - m_B
+ g_{\sigma B} \sigma - g_{\omega B} \gamma_\mu \omega^\mu
- g_{\rho B}
\gamma_\mu{\mbox{\boldmath $\tau$}}_B \cdot
{\mbox{\boldmath $\rho$}}^\mu \right)\Psi_B - \frac{1}{2} \partial_\mu \sigma\partial^\mu \sigma
-\frac{1}{2} m_\sigma^2 \sigma^2 - \frac{1}{3}b m \left(g_\sigma \sigma\right)^3 
- \frac{1}{4}c  \left(g_\sigma \sigma\right)^4 \nonumber\\
&& -\frac{1}{4} \Omega_{\mu\nu}\Omega^{\mu\nu}
-\frac{1}{2}m_\omega^2 \omega_\mu \omega^\mu
- \frac{1}{4}{\mbox {\boldmath $\mathrm{P}$}}_{\mu\nu} \cdot
{\mbox {\boldmath $\mathrm{P}$}}^{\mu\nu}
- \frac{1}{2}m_\rho^2 {\mbox {\boldmath $\rho$}}_\mu \cdot
{\mbox {\boldmath $\rho$}}^\mu ~,
\end{eqnarray}
\end{widetext}
where, $m_B$ is the baryon mass. We use the nucleon mass $m$ as the average of the baryon masses. The Dirac effective mass $m_*$ has been defined as $m_* = m- g_\sigma \sigma$. The $\sigma$, $\omega$ and $\rho$ mesons represent the scalar, vector and vector-isovector interactions, respectively. ${\mbox{\boldmath $\tau$}}_B$ is the isospin operator. $\Omega_{\mu \nu}$ and ${\mbox{\boldmath $\mathrm{P}$}}_{\mu \nu}$ are the field tensors for $\omega$ and $\rho$ mesons respectively. For the two-fluid system, we choose a frame in such a way that the neutrons have zero spatial momentum and the proton momentum has a boost along the z-direction as $k_p^\mu = \left(k_0, 0, 0, K\right)$. We follow the procedure as described in Refs. \cite{Kheto2014,Kheto2015} to solve the meson field equations and numerically evaluate the master function $\Lambda$, generalized pressure $\Psi$ etc. in the limit $K \rightarrow 0$.

We consider a normal fluid envelope around the superfluid core of the star to account for the behavior of the low density region of a NS. We assume this region to be free of superfluid neutrons. This assumption does not affect the macroscopic structure of the star. To describe the matter in this region, we employ the EOS for the inner crust calculated by Grill \textit{et al.} \cite{grill2014}. We smoothly join the EOS by keeping the pressure continuous from the two-fluid region to the envelope. We also use the DH EOS \cite{haensel2007} for the outer part of the envelope.

\subsection{Equilibrium configuration}
 \label{equil}

We take the background metric of the star to be static and spherically symmetric. Under such assumptions, the metric can be written in the Schwarzschild form as follows,

\begin{equation}
\label{equilibrium metric}
ds_0^2 = g^{(0)}_{\alpha\beta} dx^{\alpha} dx^{\beta}= -e^{\nu(r)} dt^2 + e^{\kappa(r)} dr^2 + r^2 (d\theta^2 + \sin^2\theta d\phi^2)
\end{equation}

This metric structure is valid both in the core and the envelope. Only the energy-momentum tensor changes from one region to another.

\subsubsection{Superfluid core}

In the core the energy momentum tensor will take that of an SF matter, as has been described in Eq.(\ref{SF EMT}). The two metric functions can then be evaluated from the Einstein's equations as follows,
\begin{eqnarray}
\kappa^{\prime} &=& \frac{1-e^{\kappa}}{r} - 8\pi r e^{\kappa}\Lambda |_0, \nonumber\\
\nu^{\prime} &=& -\frac{1-e^{\kappa}}{r}  + 8\pi r e^{\kappa}\Psi |_0,
\label{tov_munu}
\end{eqnarray}
By the following equations the radial profiles for $n(r)$ and $p(r)$ are determined,\cite{Comer1999},
\begin{eqnarray}
{\cal A}^0_0 |_0 p' + {\cal B}^0_0 |_0 n' + \frac{1}{2} \mu|_0\nu ' &=& 0, \nonumber\\
{\cal C}^0_0 |_0 p' + {\cal A}^0_0 |_0 n' + \frac{1}{2} \chi|_0\nu ' &=& 0,
\label{tov_np}
\end{eqnarray}
where, 
\begin{equation}
\begin{split}
{\cal A}^0_0 &= {\cal A} +2\frac{\partial {\cal B}}{\partial p^2} np + 2\frac{\partial {\cal A}}{\partial n^2}n^2 + 2\frac{\partial {\cal A}}{\partial p^2}p^2 + \frac{\partial {\cal A}}{\partial x^2} np,\\
{\cal B}^0_0 &= {\cal B} +2\frac{\partial {\cal B}}{\partial n^2}n^2 + 4\frac{\partial {\cal A}}{\partial n^2}np + \frac{\partial {\cal A}}{\partial x^2}p^2,\\
{\cal C}^0_0 &= {\cal C} +2\frac{\partial {\cal C}}{\partial p^2}p^2 + 4\frac{\partial {\cal A}}{\partial p^2}np + \frac{\partial {\cal A}}{\partial x^2}n^2.
\end{split}
\end{equation}
The two Fermi wave numbers $k_n$ and $k_p$ are the variables that are more appropriate for the RMF calculations. Thus, we substitute the number densities with the Fermi wave numbers using $n = \frac{k_n^3}{3\pi^2}$ and $p = \frac{k_p^3}{3\pi^2}$, and solve for $k_n$ and $k_p$ instead. We determine the Dirac effective mass $m_*|_0(k_n,k_p)$ using the method discussed in \cite{Comer2003}. The transcendental algebraic relation in Eq. \ref{m star eqn} is turned into a differential equation using,
\begin{equation}
m_*' |_0 = \frac{\partial m_*}{\partial k_n}\bigg |_0 k_n' + \frac{\partial m_*}{\partial k_p}\bigg |_0 k_p',
\end{equation}
where $k_n'$ and $k_p'$ are calculated from Eq. \ref{tov_np}.
The prime in the equation represents a radial derivative and a zero subscript represents that $K\rightarrow 0$ has been taken after the partial derivatives are calculated.
We put the boundary condition at the center and the surface of the star. A non-singularity condition at the center imposes $\kappa(0) = 0$ and $\kappa'(0)$ and $\nu'(0)$ vanishes. Together with Eq. \ref{tov_np} this condition imposes $k_n'(0) = k_p'(0) = 0$.  Necessary expressions for all the matter quantities used in our calculations $\big(\Lambda|_0, \Psi|_0, \mu|_0, \chi|_0, m_*|_0 , {\cal A}|_0, {\cal B}|_0, {\cal C}|_0, {\cal A}^0_0|_0, {\cal B}^0_0|_0, {\cal C}^0_0|_0$, $\left.\frac{\partial m_*}{\partial k_n}\right|_0,\left.\frac{\partial m_*}{\partial k_p}\right|_0 \big)$ can be found in Appendix \ref{Expressions for matter variables}.

\subsubsection{Normal fluid envelope}
In the envelope the matter is modeled as one component normal fluid (NF). Therefore the energy momentum tensor can be written as,

\begin{equation}
\label{SF EMT}
T^{\mu}_{\nu} = p \delta^{\mu}_{\nu} + (\rho + p)u^{\mu}u_{\nu},
\end{equation}
where $\rho$ and $p$ are the energy density and the pressure of the fluid in the envelope, respectively. And $u^{\mu}$ is the four velocity of the fluid. 

Using this form of energy-momentum tensor equation for the two metric functions can be evaluated from the Einstein's equations as follows,
\begin{eqnarray}
\kappa^{\prime} &=& \frac{1-e^{\kappa}}{r} + 8\pi r e^{\kappa}\rho, \nonumber\\
\nu^{\prime} &=& -\frac{1-e^{\kappa}}{r}  + 8\pi r e^{\kappa} p,
\label{tov_munu crust}
\end{eqnarray}

The continuity of the metric variables at the junction of the SF core and the normal fluid envelope has been discussed in appendix \ref{Junction condition for even}. The surface of the star $r=R$ implies that the total mass of the star is,
\begin{equation}
M = -4\pi \int^{R_c}_0 dr r^2 \Lambda|_0(r) +4\pi \int_{R_c}^R dr r^2 \rho(r),
\label{tov_mass}
\end{equation}
and $\Psi|_0(R_c) = p(R_c)$ and $p(R) = 0$, where $R_c$ is the junction between the SF core and the NF envelope.

\section{Even Parity Perturbation Equations for zero frequency mode}
\label{even perturbation}

To calculate the electric type tidal Love no., perturbation of the static and spherically symmetric background needs to be calculated. For this purpose we decompose the metric as \cite{Thorne1967},

\begin{equation}
g_{\alpha\beta} = g^{(0)}_{\alpha\beta} + \delta g_{\alpha\beta},
\end{equation}
where, $g^{(0)}_{\alpha\beta}$ and $\delta g_{\alpha\beta}$ are the background and the perturbed part of the metric respectively.

We decompose the metric and the fluid perturbation on the basis of spherical harmonics $Y^m_l(\theta, \phi)$. Because of the spherical symmetry of the background we take $m=0$ without breaking any generality \cite{Chandra}. Therefore the basis is the Legendre polynomials $P_l(\theta)$. 

It is well known that the perturbation can be decomposed into two kinds of classes according to their behavior under parity transformation. In this section, we will focus only on the even parity modes. For the even parity we focus on the static perturbations. Thus, the perturbations will have no explicit time dependence. After restricting ourselves in these conditions we choose the Regge-Wheeler gauge to fix the even parity perturbation $(\delta g^{(e)}_{\alpha\beta})$in the following form \cite{Regge1957},
\begin{equation}\label{even metric}
\begin{split}
 &\sum_l diag[-e^{\nu(r)}H_{0}^{(l)}(r), e^{\kappa(r)}H_2^{(l)}(r), r^2K^{(l)}(r), r^2\sin^2\theta K^{(l)}(r)]\\
 &\times  P_l(\theta)
\end{split}
\end{equation}
where, ${(e)}$ represents even parity sector.

\subsection{Superfluid core}

It is simple to calculate the perturbation in the energy momentum tensor. It can be expressed as, $\delta T^0_0 = \delta \Lambda$ and $\delta T^i_j = \delta \Psi\delta^i_j$ . Using these in the Einstein equation and keeping only the first order of the perturbation, we can find the perturbed metric equations.

\begin{equation}
\begin{split}
\delta &G^{\theta}_{\theta}-\delta G^{\phi}_{\phi} = 0\\
\implies &H_0^{(l)} = -H_2^{(l)} \equiv H^{(l)}
\end{split}
\end{equation}

\begin{equation}
\begin{split}
\delta &G^{\theta}_{\theta}+\delta G^{\phi}_{\phi} = 16 \pi \delta \Psi\\
\implies & 2\delta \Psi = P_l(\theta) H^{(l)} ( \Lambda-\Psi)
\end{split}
\end{equation}

\begin{equation}
\begin{split}
\delta &G^{r}_{\theta} = 0\\
\implies & K^{(l)\prime} + H^{(l)\prime} + H^{(l)} \nu^{\prime} = 0,
\end{split}
\end{equation}
where $\delta  G^r_r = 8\pi \delta T^r_r$ implies,
\begin{equation}
\begin{split}
 K^{(l)} = &\frac{-r^2 \nu^{\prime} H^{(l)\prime}}{(l^2+l-2)e^{\kappa}}\\
&+ \frac{H^{(l)}\{2-r^2\nu^{\prime 2} + e^{\kappa} (8\pi r^2 (\Psi - \Lambda)-l(l+1))\}}{(l^2+l-2)e^{\kappa}}
\end{split}
\end{equation}

From the linearized Euler equation we find,
\begin{equation}
\label{perturbed Euler}
\partial_t\delta\mu_i = \partial_i\delta\mu_t,\,\,\,\,\partial_t\delta\chi_i = \partial_i\delta\chi_t.
\end{equation}

Staticity implies $\delta\mu_0 = \delta\chi_0 = 0$. From [\ref{perturbed Euler}] it is straightforward to show that,
\begin{equation}
\label{mu-chi equation}
\begin{split}
&\delta\mu_0 = (A^0_0\delta p + B^0_0 \delta n)u^0\delta g_{00}+u^0\frac{\mu}{2}\delta g_{00}\\
&\delta\chi_0 = (A^0_0\delta n + C^0_0 \delta p)u^0\delta g_{00}+u^0\frac{\chi}{2}\delta g_{00}.
\end{split}
\end{equation}

Using Eqs.(\ref{even metric}), (\ref{perturbed Euler}) and (\ref{mu-chi equation}) we find,
\begin{equation}
\begin{split}
&\delta n = \frac{(\chi A^0_0-\mu C^0_0)}{(B^0_0C^0_0-A^{02}_0)}\frac{H^{(l)}P_l(\theta)}{2}\\
&\delta p = \frac{(\mu A^0_0-\chi B^0_0)}{(B^0_0C^0_0-A^{02}_0)}\frac{H^{(l)}P_l(\theta)}{2}
\end{split}
\end{equation}

$\Lambda$ is a function of $n^2, p^2$ and $x^2$. Therefore,

\begin{equation}
\begin{split}
\delta \Lambda &= \frac{\partial \Lambda}{\partial x^2}\delta x^2+\frac{\partial \Lambda}{\partial p^2}\delta p^2+\frac{\partial \Lambda}{\partial n^2}\delta n^2\\
&=-[(An+Cp)\delta p+(Ap+Bn)\delta n]\\
&= -g\frac{H^{(l)}}{2}P_l(\theta).
\end{split}
\end{equation}

 where,
\begin{equation}
\begin{split}
g = \frac{\mu^2 C^0_0+\chi^2 B^0_0-2\mu\chi A^0_0}{ A^{02}_0-B^0_0 C^0_0}
\end{split}
\end{equation}

We use the following Einstein equation along with the expression of $\delta \Lambda$ to calculate the final perturbation equation,

\begin{equation}
\delta G^{t}_{t} - \delta G^{r}_{r}= -4\pi g H^{(l)} P_l(\theta)+4P_l(\theta)\pi  H^{(l)} (\Psi - \Lambda)
\end{equation}

After some calculation this reduces to,

\begin{equation}
\label{final H equation}
\begin{split}
&H^{(l)\prime\prime}+H^{(l)\prime} \bigg[ 4 \pi  re^{\kappa } (\Lambda +\Psi )+\frac{e^{\kappa  }+1}{r}\bigg]\\
&+H^{(l)}\bigg[ 4 \pi e^{\kappa  }(-5 \Lambda +9 \Psi -g)-\nu '^2- \frac{l(l+1)e^{\kappa }}{r^2}\bigg]=0
\end{split}
\end{equation}

This is the central equation for the determination of the
electric type tidal Love numbers. Note that Eq.(\ref{final H equation})
contains the coefficients $A_{\mu\nu}$, $B_{\mu\nu}$ and $C_{\mu\nu}$ which have been evaluated in the equilibrium configuration. The main difference between Eq. (\ref{final H equation}) and its non-superfluid single fluid counterpart Eq. (15) in Ref. \cite{Hinderer2008} is as follows. In the case of the normal fluid, it is assumed that the fluid is barotropic in nature. Therefore, it is possible to write $\delta \rho = \frac{d\rho}{dp}\delta p$ and substitute it in the perturbed Einstein equations. For any multi-fluid scenarios, this
assumption is incorrect, in general. For this reason, we calculate $\delta \Lambda$ explicitly  with respect to the fluid and the perturbed metric variables. Because of this, the final equation of even parity perturbation gets modified and so does the response to the perturbation subsequently.

\subsection{Normal fluid envelope}

We model the low density region as the one component normal fluid matter. Hence, it is simple to calculate the perturbation in the energy momentum tensor of the fluid. It can be expressed as, $\delta T^0_0 = -\delta \rho = -\frac{d\rho}{dp} \delta p$ and $\delta T^i_j = \delta p\delta^i_j$ . Using these in the Einstein equation and keeping only the first order of the perturbation, perturbed metric equations have been found in several works \cite{Flanagan2008, Damour2009}. The equation is as follows,

\begin{equation}
\label{H equation in crust}
\begin{split}
&H^{(l)\prime\prime}+H^{(l)\prime} \bigg[ 4 \pi  re^{\kappa } (-\rho +p )+\frac{e^{\kappa }+1}{r}\bigg]\\
&+H^{(l)}\bigg[ 4 \pi e^{\kappa}(5 \rho +9 p +\frac{\rho + p}{dp/d\rho})-\nu '^2- \frac{l(l+1)e^{\kappa}}{r^2}\bigg]=0
\end{split}
\end{equation}

We take the initial condition for $H^{(l)}$ in the normal fluid region to be the value of the $H^{(l)}$ at the junction, found by solving  Eq.(\ref{final H equation}). Then the solution of Eq.(\ref{H equation in crust}) gives the perturbation for the entire star.

\section{Odd Parity Perturbation Equations for zero frequency mode}
\label{odd perturbation}

In this section we discuss the odd parity perturbation of the Einstein equation that will lead to the calculation of the magnetic type Love number. The zero frequency limit in the odd parity sector is discontinuous, as has been discussed in Ref. \cite{Pani_2018}. Keeping this in mind we take a time dependent perturbation of the metric and finally in the end we take the zero frequency limit carefully. After choosing the Regge-Wheeler gauge the metric perturbation $(\delta g^{(o)}_{\alpha\beta})$ can be written as follows,

\begin{equation}
    \begin{split}
    \delta g^{(o)}_{\alpha\beta}dx^{\alpha}dx^{\beta} = \sum_l &2(h_0^{(l)}(r,t)  dt d\phi + h_1^{(l)}(r,t)  dr d\phi)\\
&\times \sin\theta \partial_{\theta}P_l(\theta).
\end{split}
\end{equation}
where $(o)$ represents odd parity.

\subsection{Superfluid core}

For the odd parity modes $\delta n =0= \delta p$ where, $\delta p$ and $\delta n$ are the perturbed number density of the the proton and the neutron, respectively. If the perturbed velocity of the neutron and the proton are, respectively, $\delta u_{\mu}$ and $\delta v_{\mu}$ then only non-zero components can be written as \cite{Comer1999},

\begin{equation}
    \begin{split}
        \delta u_{\phi} &= e^{-\nu/2}\dot{U_n}(r,t)\sin\theta\frac{\partial P_l}{\partial\theta},\\
        \delta v_{\phi} &= e^{-\nu/2}\dot{U_p}(r,t)\sin\theta\frac{\partial P_l}{\partial\theta}.
    \end{split}
\end{equation}
where $U_n$ and $U_p$ are two arbitrary functions yet to be determined and $P_l$ is the Legendre polynomial.

Using the form of the velocity and metric perturbation in the Einstein equation, equation for the perturbations can be found. The equations relevant for our works are as follows:

\begin{equation}\label{odd eq1}
        \begin{split}
            &\bigg(\frac{1}{e^{\kappa}}\big[\frac{\nu'-\kappa'}{2r}+\frac{1}{r^2}\big]-\frac{l(l+1)}{2r^2}\bigg)h_1^{(l)}-\frac{1}{2e^{\nu}}\ddot{h_1}^{(l)} \\
            &+ \frac{1}{2e^{\nu}}\bigg(\dot{h_0}^{(l)\prime}-\frac{2}{r}\dot{h_0}^{(l)}\bigg) = 4\pi (\Psi+\Lambda)h_1^{(l)}
        \end{split}
\end{equation}

\begin{equation}\label{odd eq2}
    \frac{1}{e^{\nu}}\dot{h_0}^{(l)}-\frac{1}{e^{\kappa}}\bigg(h_1^{(l)\prime}+\frac{\nu'-\kappa'}{2}h_1^{(l)}\bigg)=0.
\end{equation}
A new master function is defined as, $\psi = e^{(\nu-\kappa)/2}\frac{h_1}{r}$ \cite{Pani_2018}. Equation (\ref{odd eq2}) now can be written as,

\begin{equation}
    \dot{h_0}^{(l)} = e^{(\nu-\kappa)/2}(\psi^{(l)} r)'.
\end{equation}

We take the time dependence of each mode as $e^{i\omega t}$. Putting everything together Eq.(\ref{odd eq1}) can be written as,

\begin{equation}
\label{omega psi equation}
\begin{split}
    &\psi^{(l)\prime\prime} + \frac{\psi^{(l)\prime} e^{\kappa}}{r^2}\bigg[2M(r) + 4\pi r^3(\Psi + \Lambda)\bigg]\\
&-e^{\kappa}\psi^{(l)} \bigg[-e^{-\nu}\omega^2-\frac{6M(r)}{r^3}-4\pi(\Psi+\Lambda)+\frac{l(l+1)}{r^2}\bigg] = 0.
\end{split}
\end{equation}
 After taking the $\omega\rightarrow 0$ limit the zero frequency equation takes the following form,

\begin{equation}
\label{final psi equation}
\begin{split}
    &\psi^{(l)\prime\prime} + \frac{\psi^{(l)\prime} e^{\kappa}}{r^2}\bigg[2M(r) + 4\pi r^3(\Psi + \Lambda)\bigg]\\
&-e^{\kappa}\psi^{(l)} \bigg[-\frac{6M(r)}{r^3}-4\pi(\Psi+\Lambda)+\frac{l(l+1)}{r^2}\bigg] = 0
\end{split}
\end{equation}

This is the central equation for the determination of the
magnetic type tidal Love numbers. Note that Eq.(\ref{final psi equation})
does not depend on the coefficients $A_{\mu\nu}$, $B_{\mu\nu}$ and $C_{\mu\nu}$ explicitly. But the effect of the SF nature enters through the dependence of $\Lambda$ on $x^2$. Because of this, the values of the magnetic Love numbers get modified even though the final equation of odd parity perturbation looks similar to the ones in \cite{Pani_2018, Damour2009}.

\subsection{Normal fluid envelope}

Details of the odd parity equations for normal fluid can be found in Ref.\cite{Pani_2018}. The final equation is as follows,

\begin{equation}
\label{psi equation in crust}
\begin{split}
    &\psi^{(l)\prime\prime} + \frac{\psi^{(l)\prime} e^{\kappa}}{r^2}\bigg[2M(r) + 4\pi r^3(p - \rho)\bigg]\\
    &-e^{\kappa}\psi^{(l)} \bigg[-\frac{6M(r)}{r^3}-4\pi(p-\rho)+\frac{l(l+1)}{r^2}\bigg] = 0
\end{split}
\end{equation}

We take the initial condition for $\psi^{(l)}$ in the normal fluid region to be the value of the $\psi^{(l)}$ at the junction, found by solving  Eq.(\ref{final psi equation}). Then the numerical solution of Eq.(\ref{psi equation in crust}) gives the solution for odd mode perturbation for the entire star.

\section{Calculation of the tidal Love numbers}
\label{Love}

\subsection{Electric type Love numbers}

To calculate the tidal deformability, we solve Eq. \ref{final H equation} numerically inside the NS up to the junction between the SF core and the NF envelope. Using the junction conditions described in Appendix [\ref{Junction condition for even}] we find the initial condition of $H^{(l)}$ in the envelope. This initial condition has been used to numerically evolve Eq.(\ref{H equation in crust}) up to the surface of the NS. After that the tidal Love numbers are calculated by matching the numerical value of $H^{(l)}$ found by integration with the external solution of the same equation on the surface of the star. Extensive discussion on this can be found in Refs.\cite{Hinderer2008,Binnington2009,Damour2009}.   Here we focus only on the initial conditions. We integrate Eq. \ref{final H equation} for metric perturbation in core $H^{(l)}$ radially outward from the center using the profiles of the background quantities calculated from TOV equations. For numerical purposes, instead of starting from $r=0$, we use a very small cutoff radius $(r = r_0 = 10^{-6})$. The initial condition for Eq. \ref{final H equation} around the regular singular point $r=0$ can be taken to be $H^{(l)}(r) \sim \bar{h}r^{l}$, with $\bar{h}$ some arbitrary constant. Since this equation is homogeneous and the tidal deformability depends explicitly on the value of $y^{even(l)} ~(= \frac{r H^{(l)\prime}}{H^{(l)}})$ at the surface, the scaling constant $\bar{h}$ does not hold any relevance. Therefore, we can choose the starting value for the metric variable as, $H^{(l)}(r_0) = r_0^{l}$ and $H'(r_0) = lr_0^{l-1}$.

The deformability is expressed in terms of $y^{even(l)}$, found by solving Eq.(\ref{H equation in crust}) in the envelope, and the compactness $C= \frac{M}{R}$, by matching the internal and external value of $H^{(l)}$ at the surface. The tidal Love numbers $k_2$ and $k_3$ then take the following functional form \cite{Hinderer2008,Binnington2009,Damour2009},
\begin{widetext}
\begin{equation}
\label{expr_k2}
\begin{split}
k_2 &= \frac{8}{5}(1-2C)^2C^5\big[2C(y^{(2)}-1)-y^{(2)}+2\big]\bigg[2C(4(y^{(2)}+1)C^4 + (6y^{(2)}-4)C^3+(26-22y^{(2)})C^2\\
&+3(5y^{(2)}-8)C-3y^{(2)}+6)-3(1-2C)^2(2C(y^{(2)}-1)-y^{(2)}+2)\log(\frac{1}{1-2C})\bigg]^{-1}.
\end{split}
\end{equation}
\end{widetext}

\begin{widetext}
\begin{equation}
\label{expr_k3}
\begin{split}
k_3 &= \frac{8}{7}(1-2C)^2C^7\big[2C^2(y^{(3)}-1)-3(y^{(3)}-2)C+y^{(3)}-3\big]\bigg[2C\big\{4(y^{(3)}+1)C^5 + 2(9y^{(3)}-2)C^4\\
&-20(7y^{(3)}-9)C^3+ 5(37y^{(3)}-72)C^2-45(2y^{(3)}-5)C+15(y^{(3)}-3)\big\}-15(1-2C)^2(2C^2(y^{(3)}-1)\\
&-3C(y^{(3)}-2)+y^{(3)}-3)\log(\frac{1}{1-2C})\bigg]^{-1}.
\end{split}
\end{equation}
\end{widetext}

The expression for dimensionless deformability can be found from Damour {\it et al}. to be \cite{Damour2009}, 

\begin{equation}
    \Lambda^{electric}_l =\frac{2}{(2l-1)!!} C^{-(2l+1)}k_l.
\end{equation}

Since the information of the fluid enters through $y^{(l)}|_{r=R}$ and $C$, these expressions of $k_2$ and $k_3$ are similar to the one fluid formalism. Two-fluid formalism does not change the external solution. It only changes the internal equation of $H^{(l)}$, resulting in a different value of $y^{(l)}|_{r=R}$, leading to the change in the value of  $k_l$ but not their expressions.

\begin{figure}
\includegraphics[width=9.6cm]{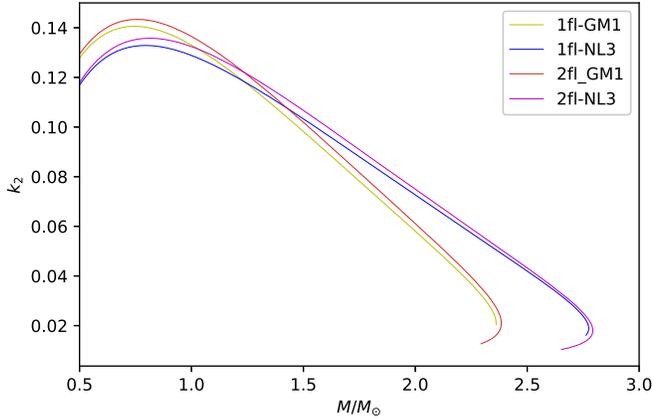}
\caption{$l=2$ electric type Love number is plotted with respect to mass of the neutron star. $M_{\odot}$ is solar mass.  }
\label{electric type l2 Love_no_plots}
\end{figure}

\begin{figure}
\includegraphics[width=9.6cm]{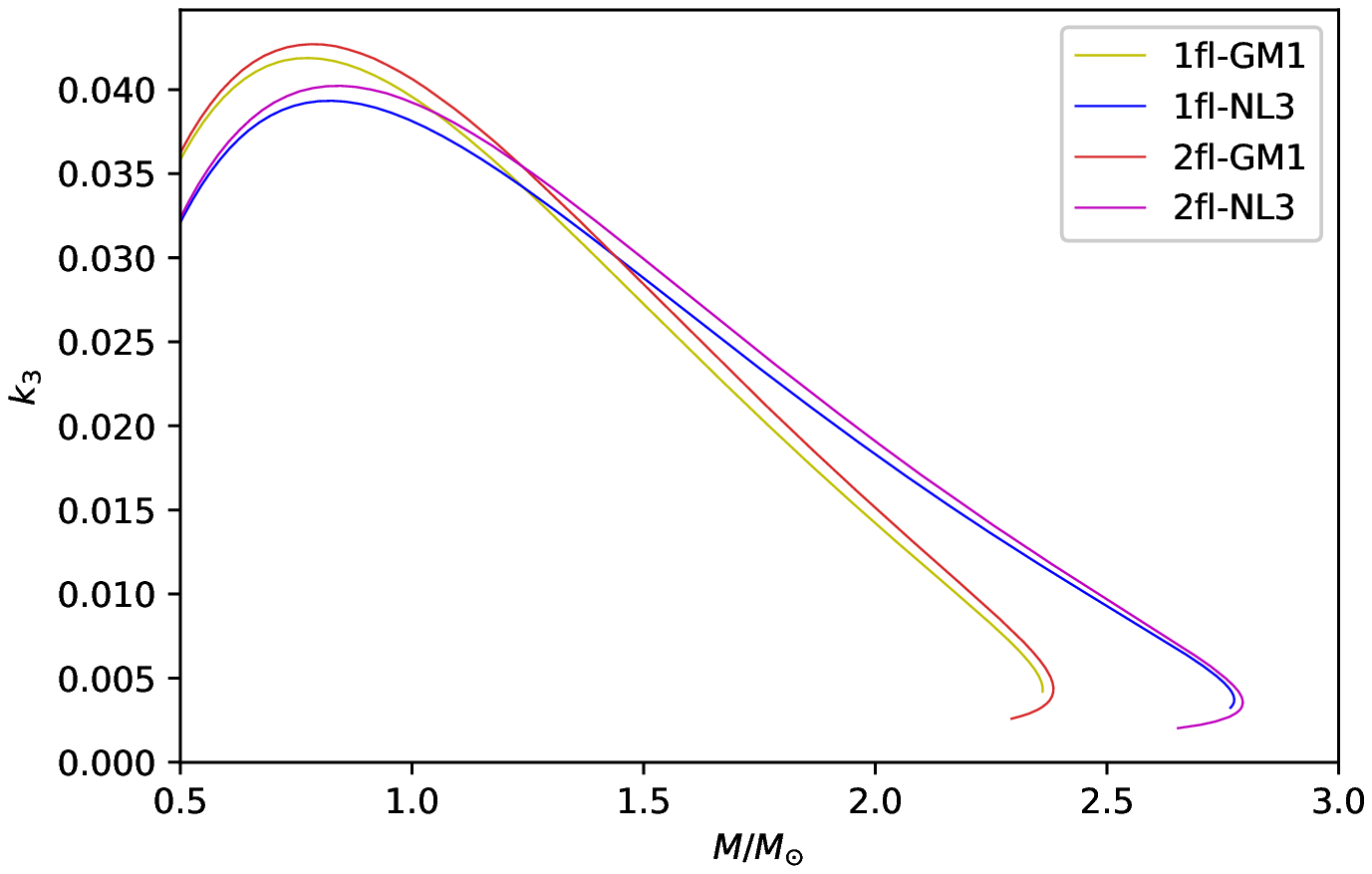}
\caption{ $l=3$ electric type Love number is plotted with respect to mass of the neutron star. $M_{\odot}$ is solar mass.}
\label{electric typ l3 Love_no_plots}
\end{figure}

\begin{figure}
\includegraphics[width=9.6cm]{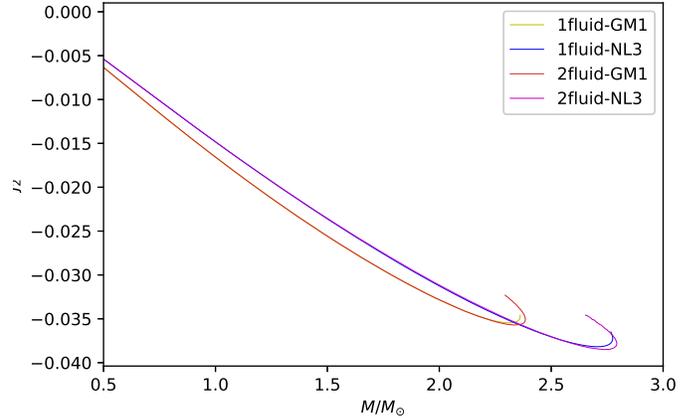}
\caption{$l=2$ magnetic type Love number is plotted with respect to mass of the neutron star. $M_{\odot}$ is solar mass.}
\label{magnetic type Love_no_plots}
\end{figure}

\subsection{Magnetic type Love numbers}

To calculate the magnetic type tidal deformability, we solve Eq. (\ref{final psi equation}) numerically inside the NS up to the junction between the SF core and the normal fluid envelope. Then using the junction conditions described in Appendix  [\ref{Junction condition for even}] we find the initial condition of $\psi^{(l)}$ in the envelope. Using this initial condition we numerically evolve Eq.(\ref{psi equation in crust}) up to the surface of the NS. The tidal Love numbers are calculated by matching the numerical value of $\psi^{(l)}$ found by integration with the external solution of the same equation on the surface of the star.  Details can be found in Ref.~\cite{Damour2009}. We will integrate Eq. \ref{final psi equation} for $\psi^{(l)}$ radially outward from the center using the profiles of the background quantities calculated from the TOV equations. Similar to the calculations of the electric-type Love number, we start from a very small cutoff radius $(r = r_0 = 10^{-6})$. The initial condition for Eq. \ref{final psi equation} near the regular singular point $r=0$ can be taken to be $\psi^{(l)}(r) \sim \bar{\psi}r^{l+1}$, with $\bar{\psi}^{(l)}$ some constant. Since, this equation is homogeneous in $\Psi^{(l)}$ and the tidal deformability depends explicitly on the value of $y^{odd(l)} ~(= \frac{r \psi^{(l)\prime}}{\psi^{(l)}})$ at the surface, the scaling constant $\bar{\psi}^{(l)}$ is not relevant. Therefore, the starting value for the metric variable can be chosen as, $\psi^{(l)}(r_0) = r_0^{l+1}$ and $\psi^{(l)\prime}(r_0) = (l+1)r_0^{l}$.

The deformability can be expressed in terms of $y^{odd(l)}$, found by solving Eq.(\ref{psi equation in crust}) in the envelope, and the compactness $C= \frac{M}{R}$, by matching the internal and external values of $\psi^{(l)}$ at the  surface. The tidal Love number $j_2$ takes the functional form \cite{Damour2009},
\begin{widetext}
\begin{equation}
\label{expr_j2}
\begin{split}
j_2 &= \frac{96}{5}(2C-1)(y-3)C^5\bigg[2C\{12(y^{(2)}+1)C^4 + 2(y^{(2)}-3)C^3+2(y^{(2)}-3)C^2\\
&+3(y^{(2)}-3)C-3y^{(2)}+9\}+3(2C-1)(y^{(2)}-3)\log(1-2C)\bigg]^{-1}.
\end{split}
\end{equation}
\end{widetext}

The expression for dimensionless deformability can be found from Damour {\it et al}. to be \cite{Damour2009}, 

\begin{equation}
    \Lambda^{magnetic}_l =\frac{(l-1)}{4(l+2)(2l-1)!!} C^{-(2l+1)}j_l.
\end{equation}

This expression of $j_2$ is similar to the one fluid formalism because the information of the fluid enters through $y^{(l)}|_{r=R}$ and $C$. The two-fluid model does not change the external solution. It changes only the internal equation of $\psi^{(l)}$, that gives us a different value of $y^{(l)}|_{r=R}$, leading to the change in the value of  $j_l$ but not its expression.

\begin{widetext}

\begin{table} 
\caption{Nucleon-meson coupling constants in 
the NL3 and GM1 sets are taken from Refs.\cite{Glendenning1991,Fattoyev2010}. The coupling constants are obtained by reproducing the saturation properties of symmetric nuclear matter as detailed in the text. All the parameters are in {\rm fm$^{2}$}, except $b$ and $c$ which are dimensionless.}

\begin{center}
\begin{tabular}{cccccc} 

\hline\hline
\hfil& 
$c_{\sigma}^2$& $c_{\omega}^2$& $c_{\rho}^2$& $b$& $c$ \\ \hline
NL3& 15.739& 10.530& 5.324& 0.002055& -0.002650 \\ \hline
GM1& 11.785&  7.148& 4.410& 0.002948& -0.001071 \\ \hline
\hline

\end{tabular}
\end{center}
\label{tab1}
\end{table}
\end{widetext}

\begin{widetext}

\begin{table} 
\caption{Values of tidal deformabilities for $1.4 M_{\odot}$}

\begin{center}
\begin{tabular}{c|c|c|c|c|c|c} 

\hline\hline
\hfil& 
$\Lambda^{el}_2$ 1-fl & $\Lambda^{el}_2$ 2-fl & $\Lambda^{el}_3$ 1-fl & $\Lambda^{el}_3$ 2-fl & $\Lambda^{mag}_2$ 1-fl & $\Lambda^{mag}_2$ 2-fl \\ \hline
NL3& 1268 & 1391 &3455  &4015.5  & -7.9 &-8.4 \\ \hline
GM1& 903&  979 & 2241 & 2440.5 & -6.2 & -6.6\\ \hline
\hline

\end{tabular}
\end{center}
\label{tab:values}
\end{table}
\end{widetext}

\begin{figure}
\includegraphics[width=9.6cm]{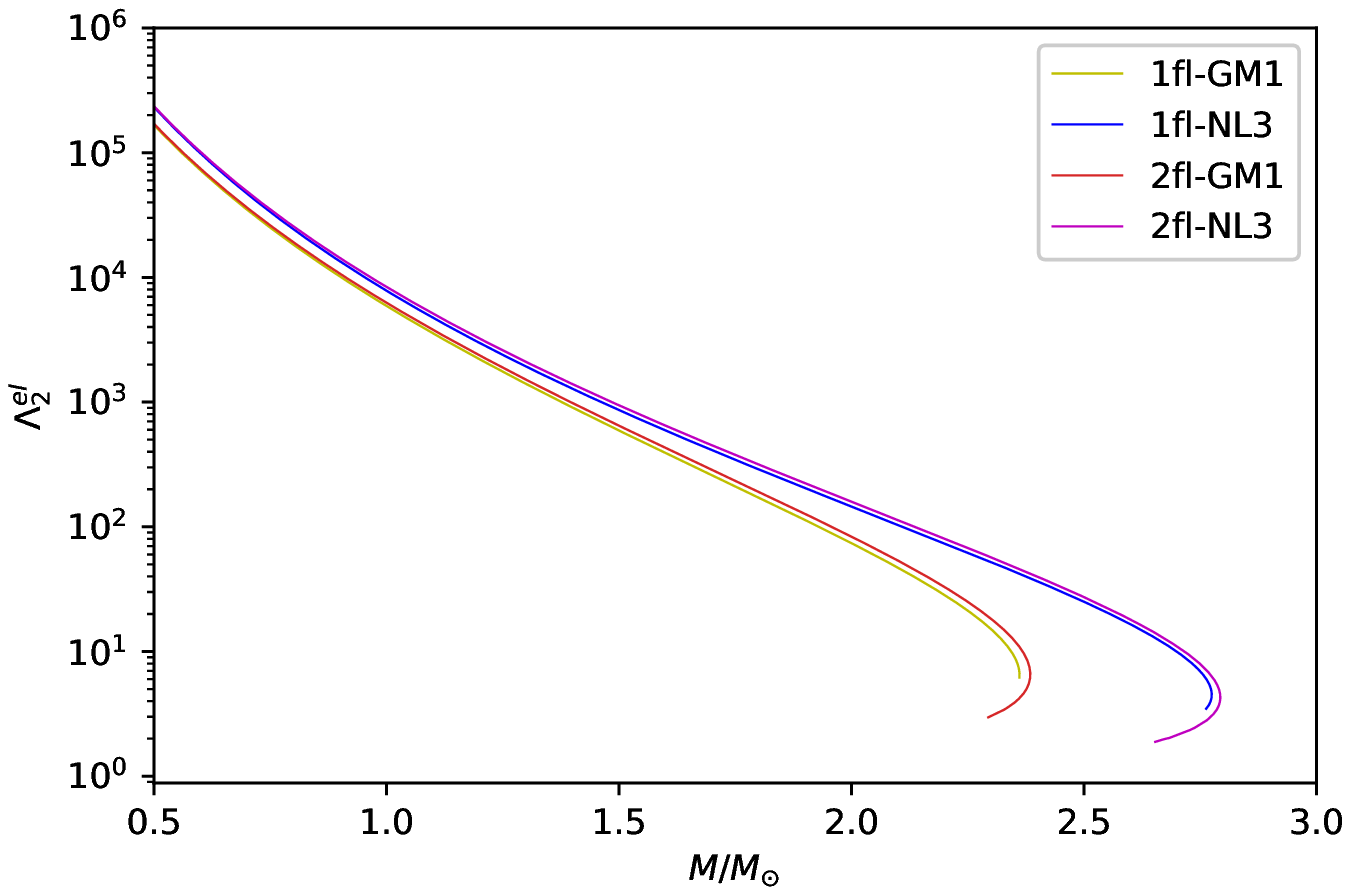}
\caption{ $l=2$ dimensionless electric type tidal deformability is plotted with respect to mass of the neutron star. $M_{\odot}$ is solar mass.}
\label{electric type l2 tidal deformability_plots}
\end{figure}

\begin{figure}
\includegraphics[width=9.6cm]{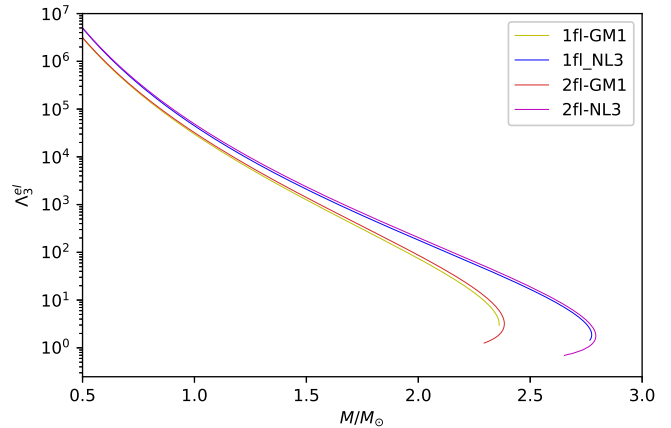}
\caption{$l=3$ dimensionless electric type tidal deformability is plotted with respect to mass of the neutron star. $M_{\odot}$ is solar mass. Dashed lines represents result in superfluid scenario and solid line represents results for one component normal fluid. Black and red colour represents NL3 and GM1 parametrization respectively.}
\label{electric type l3 tidal deformability_plots}
\end{figure}

\begin{figure}
\includegraphics[width=9.6cm]{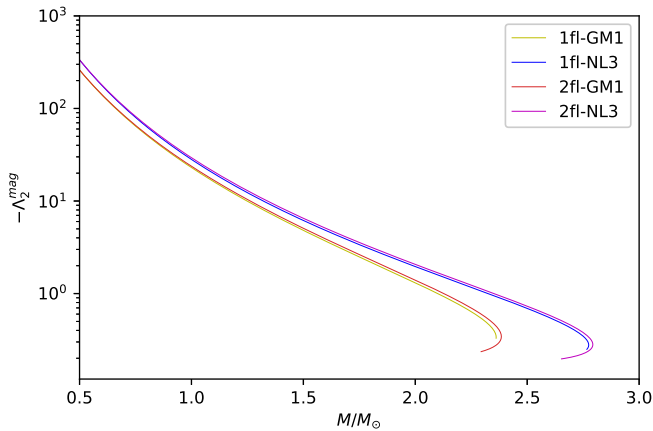}
\caption{ $l=2$ dimensionless magnetic type tidal deformability is plotted with respect to mass of the neutron star. $M_{\odot}$ is solar mass.}
\label{magnetic type l2 tidal deformability_plots}
\end{figure}

\section{Results}
\label{res}

In this section, we discuss the numerical results for tidally deformed superfluid NS. At first, we calculate the static equilibrium configurations by solving the TOV equations using realistic EOS. Since only a few calculations are available for the two-fluid system in the literature, we choose a RMF type model with scalar self-interaction terms and use NL3 and GM1 parametrizations, as in paper I. We impose $\beta$-equilibrium at the center of the star by imposing $\mu|_0 = \chi|_0$ to get a set of $k_n$, $k_p$ and $m_*$ for calculating the central number densities of the neutron and proton, energy density ($-\Lambda|_0$) and pressure ($\Psi|_0$). These quantities are used to solve Eqs. (\ref{tov_munu}), (\ref{tov_np}), (\ref{tov_munu crust}) and (\ref{tov_mass}), to find the structure of the star and to generate profiles for various background quantities for several different sets of ($k_n, k_p, m_* $) that corresponds to the different central energy densities. The maximum mass, we have found to be 2.793 $\textnormal{M}_\odot$ for NL3 and the corresponding radius being 13.34 km. Similarly, for GM1, the maximum mass is calculated to be 2.384 $\textnormal{M}_\odot$ and the corresponding radius is 12.04 km. Details of those parameter sets can be found in Table \ref{tab1}. Moreover, for NL3 and GM1 sets, the crust-core transition pressures are 0.2698 and 0.2434 MeV$/$fm$^3$ respectively. The two-fluid and the single fluid TOV integrations are smoothly joined at those pressures. Here, it is important to stress the fact that, these EOSs serve representative purposes only.

\begin{figure}
\includegraphics[width=9.6cm]{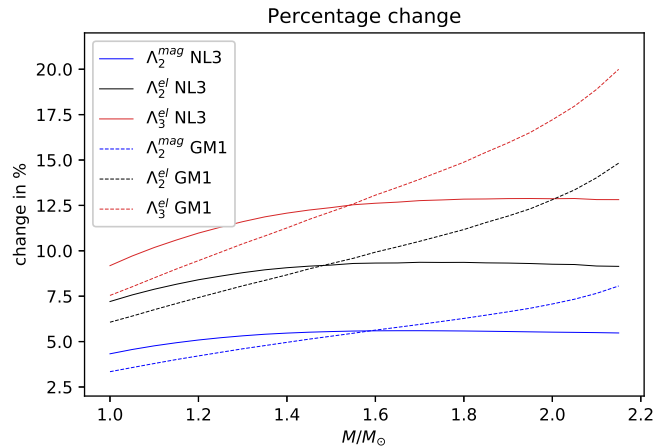}
\caption{The percentage change in dimensionless tidal deformabilities is plotted here with respect to the mass of the neutron star. $M_{\odot}$ is solar mass. }
\label{percent_plots}
\end{figure}

After getting the structure of the background, we find the numerical solution for $H^{(l)}$ for the entire star using Eqs. (\ref{final H equation}) and  (\ref{H equation in crust}) and the junction conditions described in Appendix \ref{Junction condition for even}.  Using the background profiles mentioned earlier, find $y^{even(l)}$ at the surface of the stars and calculate the electric type Love numbers using Eqs. (\ref{expr_k2}) and (\ref{expr_k3}). Similarly we find the numerical solution for $\psi^{(l)}$ for the entire star using Eqs. (\ref{final psi equation}) and (\ref{psi equation in crust}) and the junction conditions described in Appendix \ref{junction condition for odd}. Then we find $y^{odd(l)}$ at the surface of the stars and calculate the magnetic type Love number using Eq.(\ref{expr_j2}). The behavior of $k_2, k_3$ and $j_2$ w.r.t mass of the NS has been shown in the Figs.\ref{electric type l2 Love_no_plots},\ref{electric typ l3 Love_no_plots} and \ref{magnetic type Love_no_plots} respectively, along with the case of normal fluid. We plot the dimensionless tidal deformabilities in Figs. \ref{electric type l2 tidal deformability_plots}, \ref{electric type l3 tidal deformability_plots} and \ref{magnetic type l2 tidal deformability_plots} along with the normal fluid case. 
The values of the tidal deformabilities for $1.4 M_{\odot}$ is shown in Table \ref{tab:values}. We show the percentage change in Fig. \ref{percent_plots}. For all the stellar configurations, we find the tidal deformabilities of the two fluid star are larger than the normal one fluid stars.  To calculate the tidal deformabilities for the normal fluid case we used the unified EOS. As a result in both cases of NL3 and GM1, the crust is included in the calculation.

It is important to note that when we speak of the deviation of $\Lambda_{2}$ due to the superfluid nature, we bring an ambiguity in our interpretation of the observed $\Lambda_{2}$. The value of $\Lambda_{2}$ in the two-fluid calculation for a particular EOS model can be similar to the value in a single-fluid calculation for another EOS. So, we can not distinguish between the EOS and also probe the fluid nature of matter at the same time with the measurement of $\Lambda_{2}$. There are other possible degeneracies that can affect its value too \cite{Biswas2019anisotropy, Raposo2019}. In Sec. \ref{Universal relation} we discuss how this degeneracy can be broken.

\begin{figure*}[th]
\centering
\includegraphics[width=0.45\textwidth]{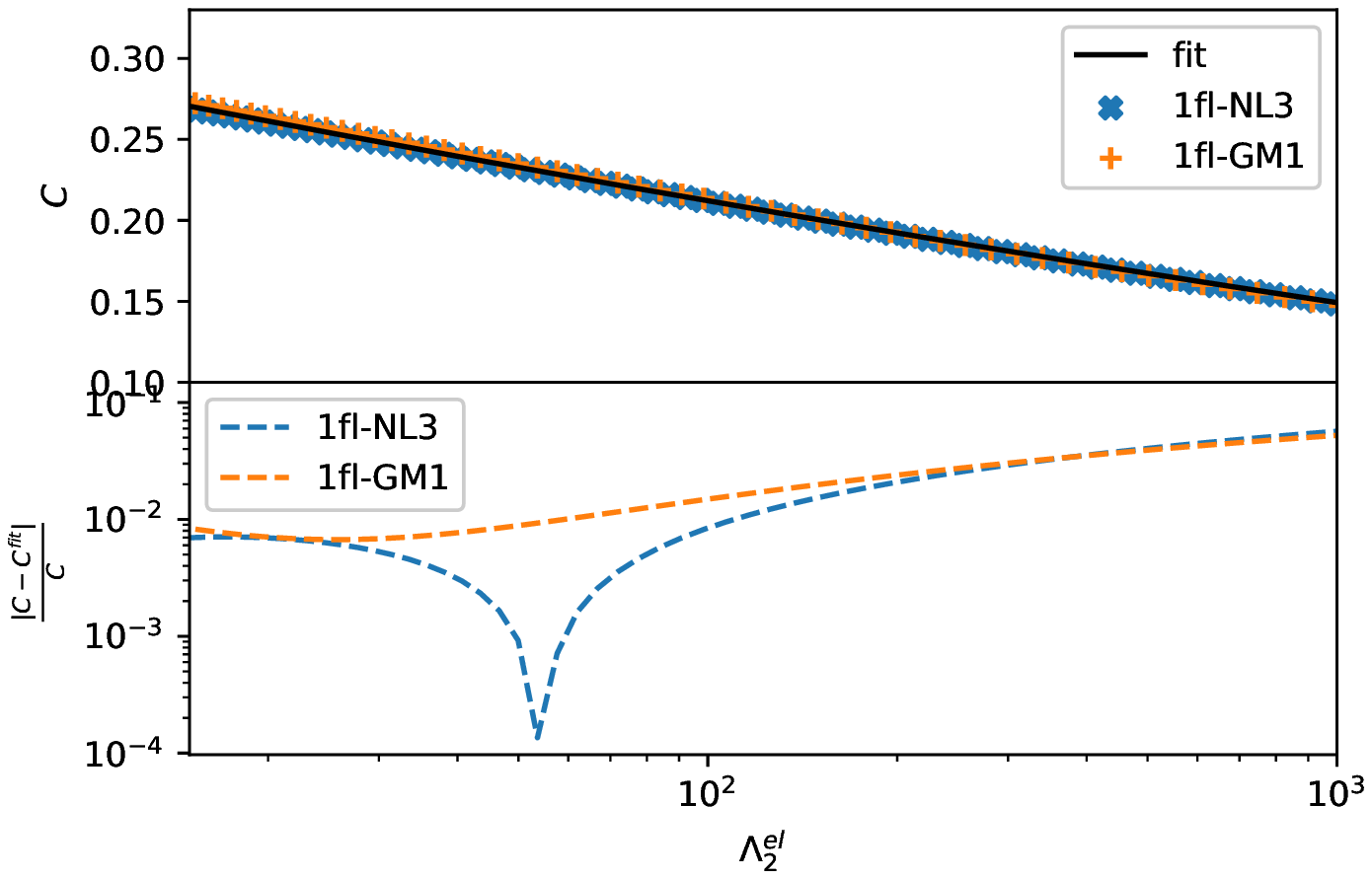}
    \includegraphics[width=0.45\textwidth]{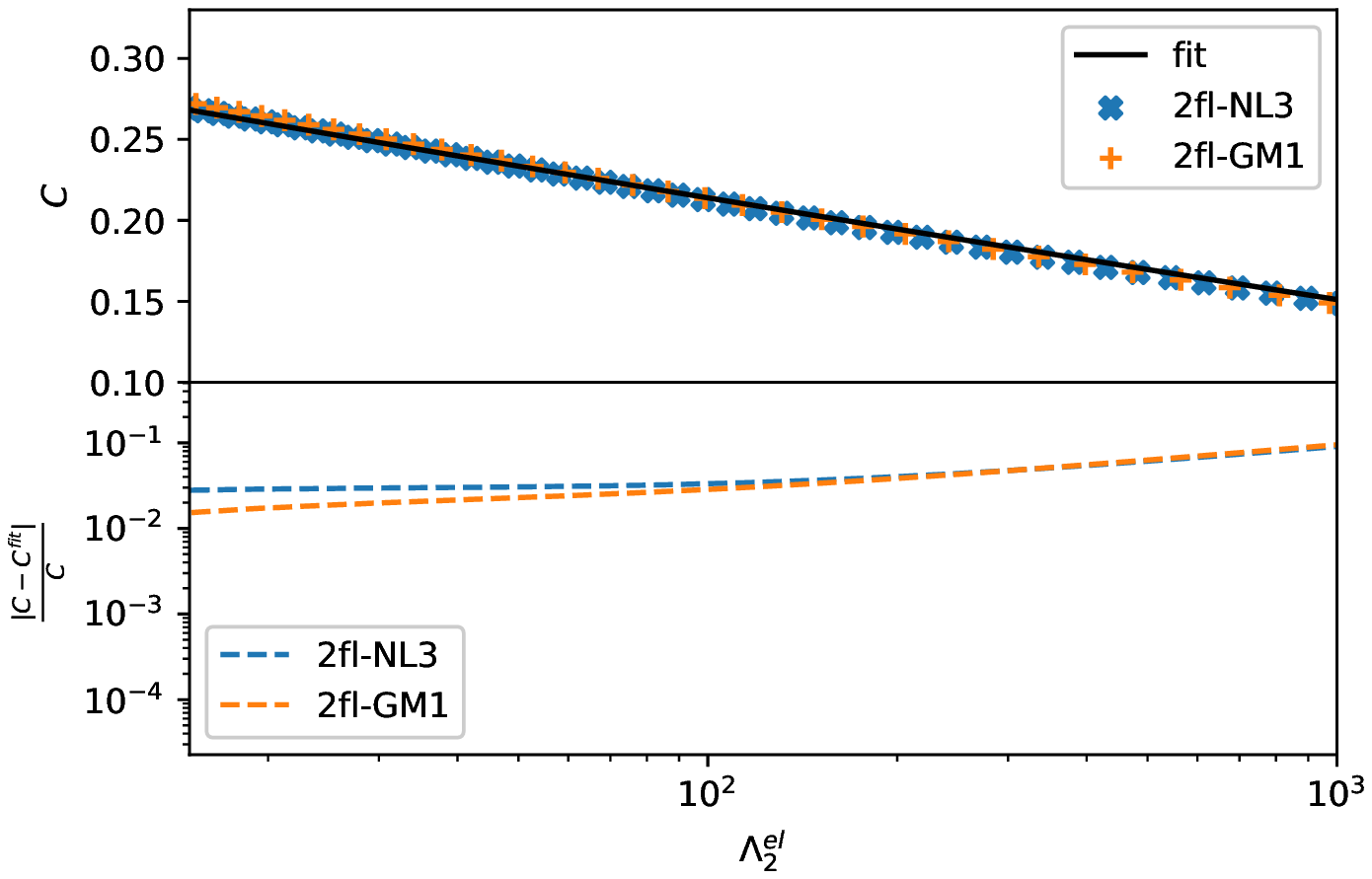}
\caption{This diagram shows the universal relation between $C$ and $
    \Lambda^{el}_2$. The upper half of the left panel shows the universality for one component normal fluid and the upper half of the right panel shows the universal relation for the two fluid system. The lower halves of both panels show errors w.r.t. fitted curves.}
\label{fig:Universal_relation_C}
\end{figure*}

\begin{figure*}[th]
\centering
\includegraphics[width=0.45\textwidth]{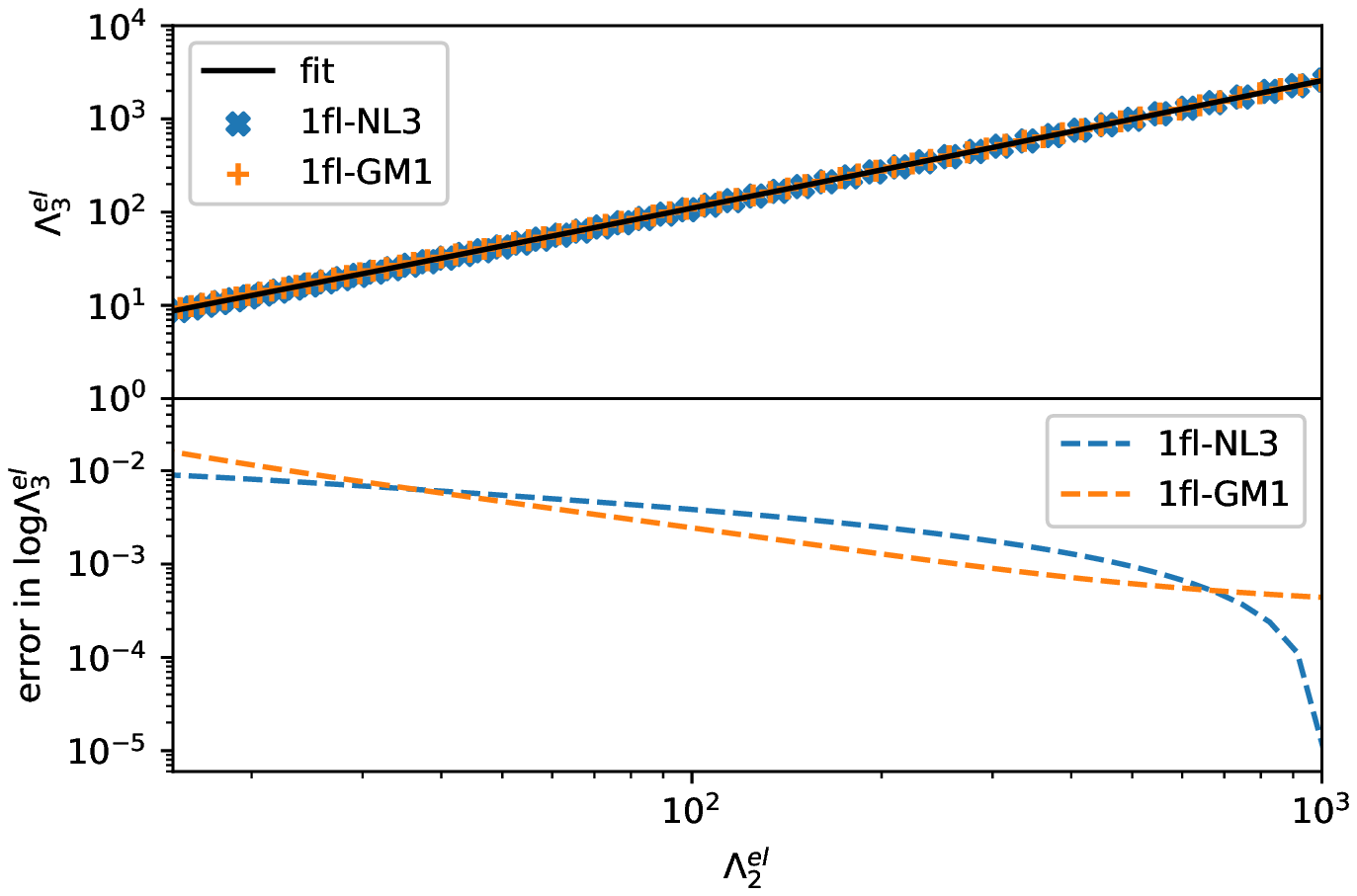}
    \includegraphics[width=0.45\textwidth]{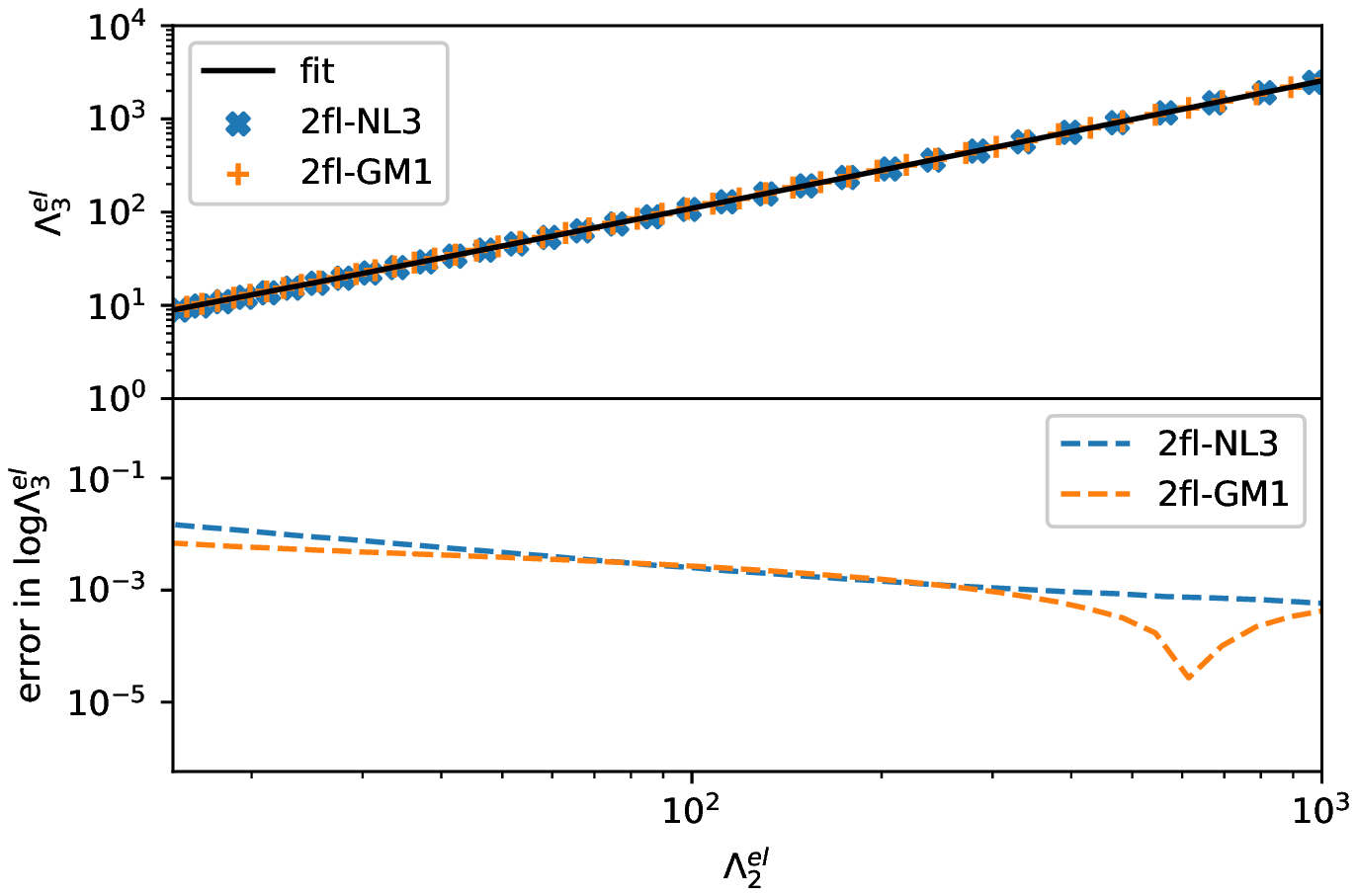}
\caption{This diagram shows the universal relation between $\Lambda^{el}_3$ and $
    \Lambda^{el}_2$. The upper half of the left panel shows the universality for one component normal fluid and the upper half of the right panel shows the universal relation for the two fluid system. The lower halve of the both panel shows error w.r.t. fitted curve.}
\label{fig:Universal_relation_lambda2_lambda_3}
\end{figure*}

\section{Universal relation}
\label{Universal relation}

In this section we fit compactness $C$, $\Lambda^{el}_3$ and $\Lambda^{mag}_2$ calculated in the previous sections against $\Lambda^{el}_2$, to test the universal relation. In Fig. \ref{fig:Universal_relation_C} we plot $C$ against $\Lambda^{el}_2$. The upper half of the left panel represents the case when the fluid has been taken to be one component normal fluid. The upper half of the right panel represents the case when the matter is modeled as a two component superfluid core and a normal fluid envelope.  For all the cases we fit them with a fitting function. The lower halves of both panels show errors w.r.t. fitted curves. In Fig. \ref{fig:Universal_relation_lambda2_lambda_3} we plot $\Lambda^{el}_3$ against $\Lambda^{el}_2$.    The upper half of the left panel represents the case when the fluid has been taken to be one component normal fluid. The upper half of the right panel represents the case when the matter is modeled as a two component superfluid core and a normal fluid envelope.  For all cases we fit them with a fitting function. The lower halves of the both panels show errors w.r.t. fitted curves. For the $C-\Lambda^{el}_2$ relation, we fit the results for both the one fluid and the SF case with the following function \cite{Yagi_2014, Maselli_2013},

\begin{equation}
C = a + b\ln \Lambda^{el}_2 + c(\ln \Lambda^{el}_2)^2.
\end{equation}

For the other cases, we used the following fitting function \cite{Yagi_2014, Maselli_2013},

\begin{equation}
y = a + b\ln \Lambda^{el}_2 + c(\ln \Lambda^{el}_2)^2 + d (\ln \Lambda^{el})^3 + f (\ln \Lambda^{el})^4.
\end{equation}

\begin{figure}
    \includegraphics[width=9.6cm]{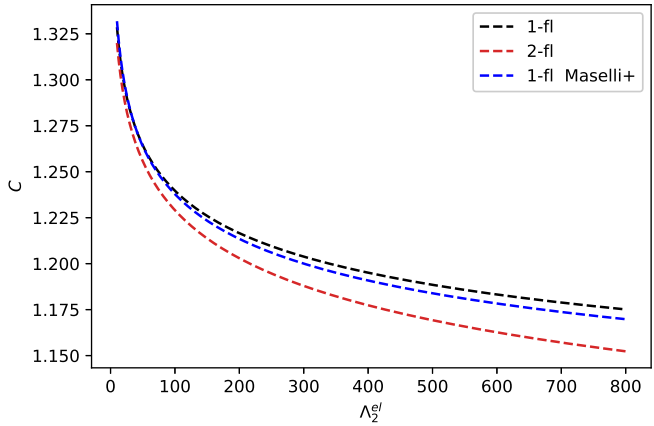}
    \caption{This diagram shows the differences between the $C-\Lambda^{el}_2$ universality curves for different fluid scenarios. For comparison we have plotted the corresponding curve using the fitting parameters from Ref. \cite{Maselli_2013}.}
    \label{fig:comparison with C}
\end{figure}

\begin{figure*}[th]
\centering
\includegraphics[width=0.45\textwidth]{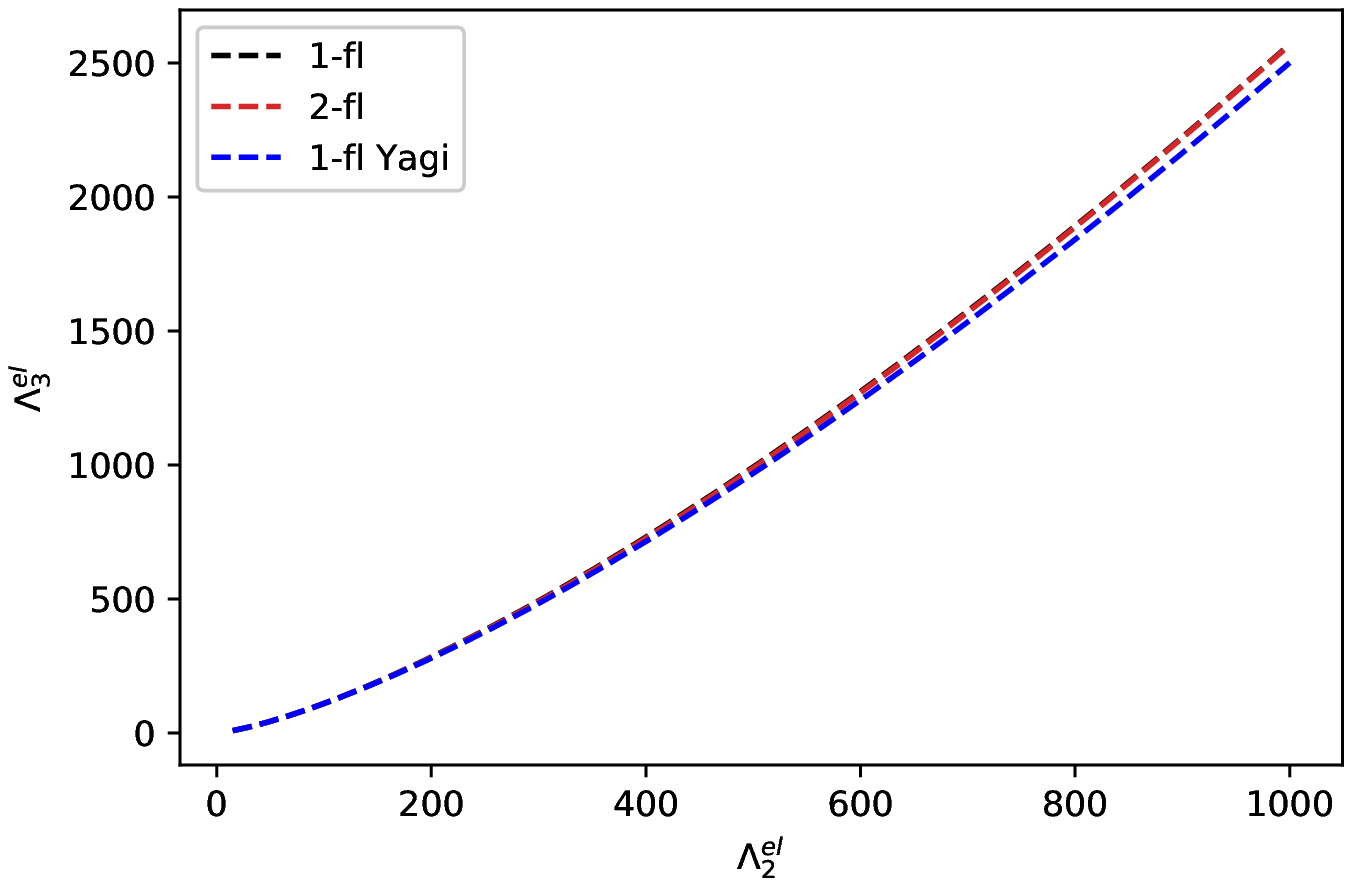}
    \includegraphics[width=0.45\textwidth]{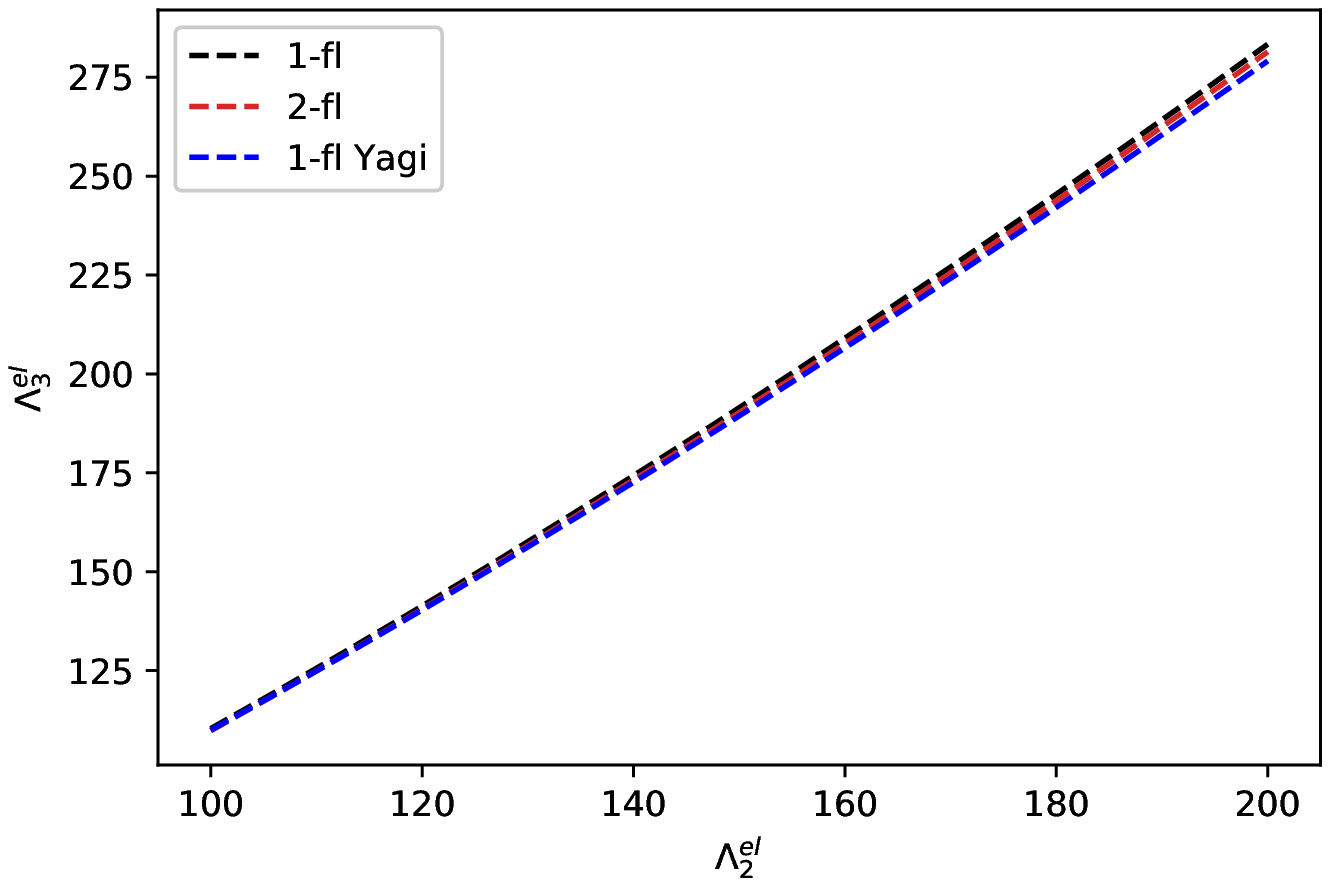}
    \includegraphics[width=0.45\textwidth]{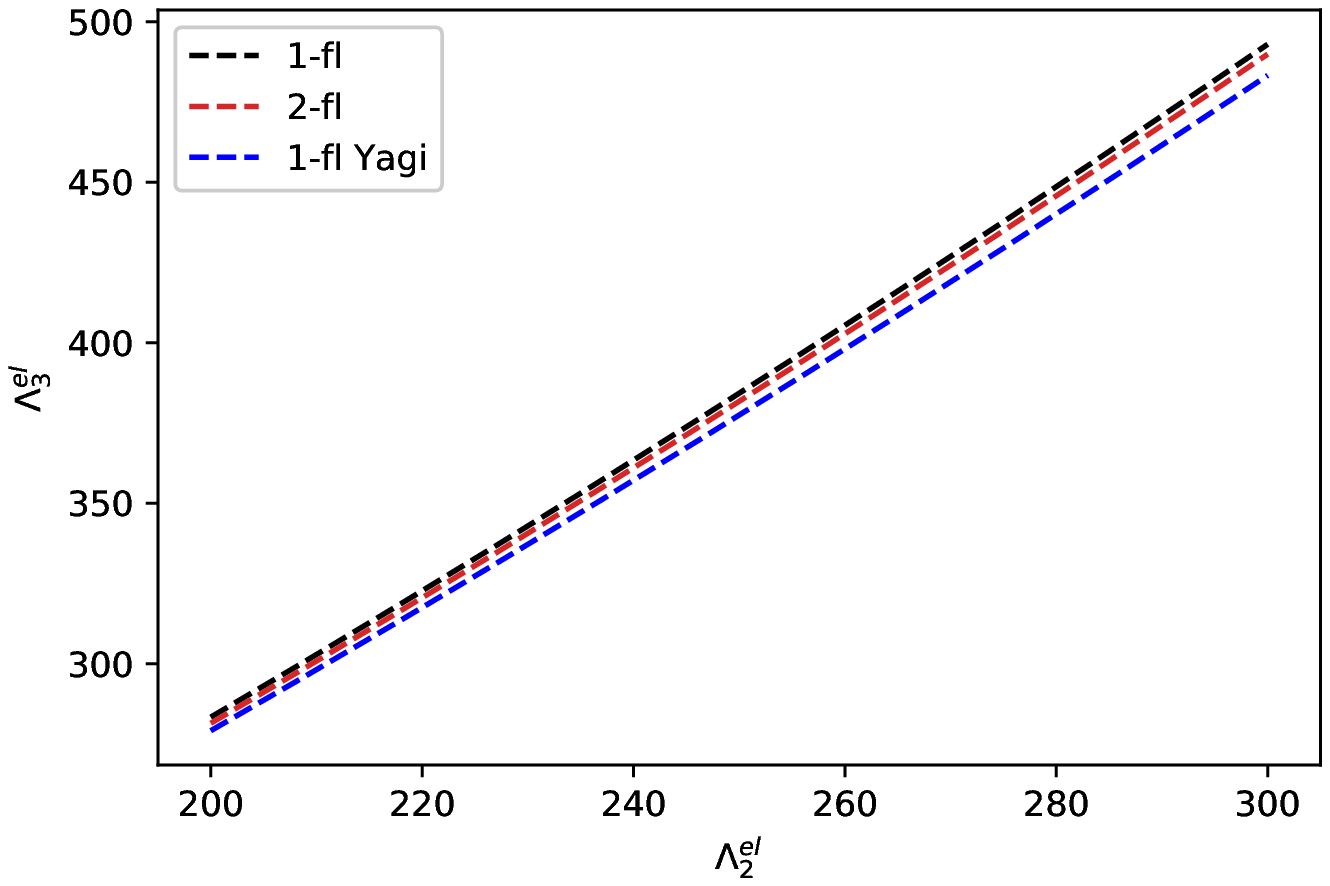}
    \includegraphics[width=0.45\textwidth]{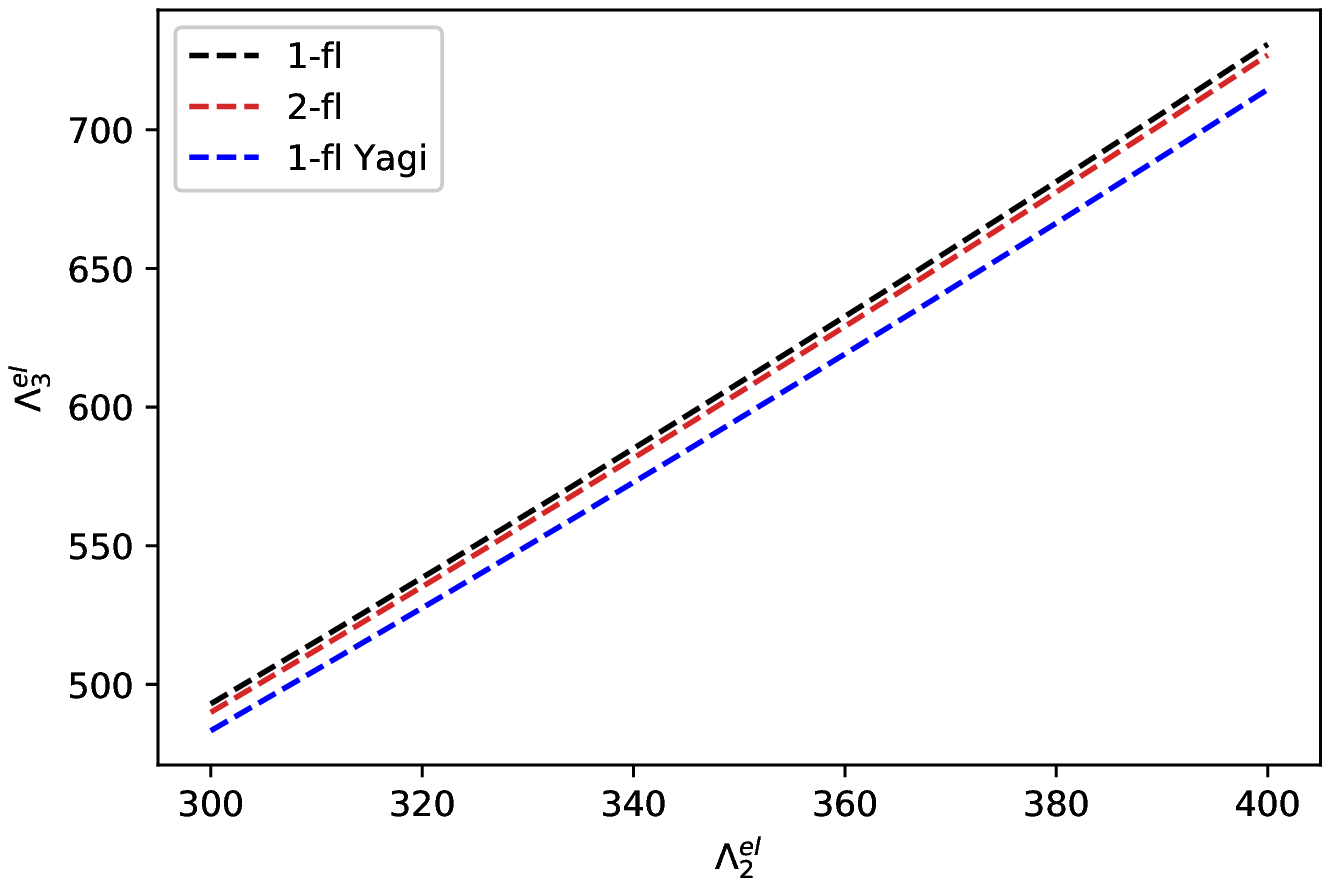}
\caption{This diagram shows the differences between the $\Lambda^{el}_3-\Lambda^{el}_2$ universality curves for different fluid scenarios. For comparison we have plotted the corresponding curves using the fitting parameters from Ref. \cite{Yagi_2014}}
\label{fig:comparison with Lambda 3}
\end{figure*}

The details of the fitted values of the parameters are described in Table \ref{table:fitting}. In Fig. \ref{fig:comparison with C} we show the differences between the $C-\Lambda^{el}_2$ universality curves for the different fluid scenario. For comparison, we have plotted the corresponding curve using the fitting parameters from Ref. \cite{Maselli_2013}, which has been named as $"1-fl$ Maselli$+"$. In Fig. \ref{fig:comparison with Lambda 3} we show the differences between the $\Lambda^{el}_3-\Lambda^{el}_2$ universality curves for different fluid scenarios. For comparison, we have plotted the corresponding curve using the fitting parameters from Ref. \cite{Yagi_2014}, which has been named as $"1-fl$ Yagi$"$.

Interestingly one fluid formalism and SF formalism both show universal behavior, even though the values of tidal deformabilities change due to the inclusion of the SF. But a crucial feature that we have found is that the fitted curve for normal fluid formalism is different from the scenario when the matter is treated using two fluid formalism. For example in Table \ref{table:fitting} it can be seen that the values of $a,b,c,d,f$ are different for two different formalisms.
This is an important observation as it opens up the possibility to probe the SF nature of matter using this deviation. From the GW data, it is possible to estimate the values of $\Lambda^{el}_2$ and $\Lambda^{el}_3$. Having an estimation of such manner it is possible to check which universal relation is more suitable for the observed values of the deformabilities. As the two different fluid natures imply different universal relations, measured values of the deformabilities will be able to distinguish between the two different universality curves. This result will help us break the degeneracy between the fluid nature and the EOS, discussed in earlier sections. It is important to note that even though the universal curves are different, they are not ``too different". Therefore it remains to be seen whether this strategy will be able to break the degeneracy with the real data.

\begin{widetext}

\begin{table} 
\caption{Estimated numerical coefficients for the fitting formula of the multipole Love relations}

\begin{center}
\begin{tabular}{cccccccc} 

\hline\hline
\hfil 
$y$ & $x$ & fluid type &$a$& $b$& $c$& $d$& $f$ \\ \hline
 $C$& $\Lambda_2^{el}$ & 1-fl & 0.364& -0.037 & 0.001 & - & - \\ 
 $C$& $\Lambda_2^{el}$ & 2-fl & 0.349 & -0.031 & 0.0 & - & - \\ 
 $\log\Lambda_3^{el}$& $\Lambda_2^{el}$ & 1-fl & -1.19050 &  1.14172 & 0.04698 & -0.00447 & 0.00017 \\ 
 $\log\Lambda_3^{el}$& $\Lambda_2^{el}$ & 2-fl & -1.09537 &  1.11772 & 0.04237 & -0.00290 & 0.00007 \\ 
 $\log-\Lambda_2^{mag}$& $\Lambda_2^{el}$ & 1-fl & -1.99175 &  0.44237 & 0.02082 & -0.00039 & -0.00000 \\ 
 $\log-\Lambda_2^{mag}$& $\Lambda_2^{el}$ & 2-fl & -1.87476 &  0.35685 & 0.04403 & -0.00317 & 0.00012 \\ \hline
\hline

\end{tabular}
\end{center}
\label{table:fitting}
\end{table}
\end{widetext}

\section{Conclusion}
Results found in the current work are very important in the context of constraining the dense matter EOS using the GW data. Values of the deformabilities for superfluid NS are higher than the normal fluid star, for a given RMF model. At present tight constraint has been put on the EOS from the BNS observation \cite{Abbott2017, Abbott2018, Nandi_2018, Zack_2019, Zack_2019_2,Bharat_2019, Luca_2019}. The results found here indicate that, more EOSs will be ruled out which are otherwise allowed if we do not consider superfluidity inside the NS. This provides us the opportunity to improve our understanding of the SF nature of the dense matter with better observational data in the future.

We find that the Love numbers are usually larger for a two-fluid system. Comer {\it et al}. \cite{Comer1999} found the existence of several superfluid oscillation modes that cannot be found otherwise in a single fluid star. This nature is very specific to the two-fluid formalism where different fluid modes can appear due to the existence of the two different types of fluid displacements. Flanagan and Hinderer discussed the fact that the tidal deformation of a star can be thought of as the sum of the deformations arising from different fluid modes that have been excited inside the star, due to the tidal perturbation \cite{Flanagan2008}. Therefore, we can say that, due to the appearance of extra fluid modes in the superfluid stars, we will get slightly larger deformations under tidal perturbation.

We argued that there is a degeneracy between the fluid nature and the EOS. Interestingly, we found that in the SF case the tidal deformabilities show a universal relation but the universal curve is different from the one fluid case. We discussed how measuring different tidal deformabilities and using universal relations can break the degeneracy between the fluid nature and the EOS.

\section*{Acknowledgements}
We thank Paolo Pani, Bhaskar Biswas, Niels Andersson, Rana Nandi and Sukanta Bose for helpful discussions. We are also grateful to H. Pais for kindly providing us with the GM1 inner crust table. This work is supported in part by the Navajbai Ratan Tata Trust. S.D. would like to thank the University Grants Commission (UGC), India, for financial support.

\appendix

\section{Junction condition}

In our current work we have modeled the NS as a superfluid core with a normal fluid envelope. Crustal physics is encoded in the current model via the normal fluid envelope. As there are two layers of fluid in our model, it is necessary to find the junction condition across the boundary. For the purpose of simplicity in this section we will use $\Psi$ as the symbol for pressure both in the SF core and in the NF envelope, while we derive the junction conditions. To calculate the junction conditions we take the level surfaces of $\Psi$.  As there are no ``delta-function like" discontinuities in $\Psi$, the first and second fundamental forms are continuous everywhere inside the star \cite{MTW}. Therefore, by imposing continuity in the first and second forms we can find the junction conditions.

The normal to the level surface of  $\Psi$ is 
\begin{equation}
    \mathcal{N}^{\mu} = \frac{g^{\mu\nu}\nabla_{\nu}\Psi}{\sqrt{\nabla_{\mu}\Psi\nabla^{\mu}\Psi}}.
\end{equation}

The induced three metric (first fundamental form) $\gamma_{\mu\nu}$ is,

\begin{equation}
    \begin{split}
        \gamma_{\mu\nu}&=\perp^{\sigma}_{\mu}\perp^{\tau}_{\nu}g_{\sigma\tau}\\
        \perp^{\sigma}_{\mu}&=\delta^{\sigma}_{\mu}-\mathcal{N}^{\sigma}\mathcal{N}_{\mu}.
    \end{split}
\end{equation}

The extrinsic curvature (second fundamental form) $\mathcal{K}_{\mu\nu}$ is defined as follows,

\begin{equation}
    \mathcal{K}_{\mu\nu} = -\perp^{\sigma}_{\mu}\perp^{\tau}_{\nu}\nabla_{(\sigma}\mathcal{N}_{\tau)},
\end{equation}

where parentheses imply symmetrization of the indices. Junction conditions will be found from the continuity of $\gamma_{\mu\nu}$ and $\mathcal{K}_{\mu\nu}$.

\subsection{Equilibrium configuration and even parity sector}
\label{Junction condition for even}

As we are mainly interested in the perturbation on the background, we write $\Psi$ as follows,

\begin{equation}
    \Psi(t,r,\theta) = \Psi_0(r) + \delta \Psi(r,\theta).
    \end{equation}

As a smooth background is constructible even in the presence of perturbation, we assume that the background and the perturbed part of $\gamma_{\mu\nu}$ and $K_{\mu\nu}$ are separately continuous at the junction. We will discuss only those components of $\gamma_{\mu\nu}$ and $K_{\mu\nu}$ that are relevant for our purpose, for more details see Ref.\cite{Andersson_2002}. First we consider the components that are useful for the even mode perturbation added to the background quantities. In the zero frequency limit the relevant quantities can be expressed as,

\begin{eqnarray}
        \gamma_{00} &= -e^{\nu} + \delta g_{00},\\
        K_{00} &= \frac{\nu'}{2}e^{\nu-\kappa/2} -\frac{1}{2e^{\kappa/2}}\delta g_{00}' + \frac{\nu'}{4}e^{\nu-3\kappa/2}\delta g_{11},\\
        K_{12} &= \frac{e^{\lambda/2}}{r}\frac{\delta \Psi_{,\theta}}{\Psi_0'}\\
        K_{22} &= -\frac{r}{e^{\kappa/2}} - e^{\kappa/2}\frac{\delta \Psi_{,\theta\theta}}{\Psi_0'} - \frac{1}{2e^{\kappa/2}}(\delta g_{22}' - \frac{r}{e^{\kappa}\delta g_{11}}).
\end{eqnarray}
    
    With $\delta \Psi(r,\theta) = \delta \Psi (r)P_l(\theta)$ these sets of equations imply,
    
    \begin{eqnarray}
         \nu(R_c) &= \tilde{\nu}(R_c)\\
         \nu'(R_c) &= \tilde{\nu}'(R_c)\\
         \kappa(R_c) &= \tilde{\kappa}(R_c)\\
         \Psi_0(R_c) &= \tilde{\Psi}_0(R_c)\\
            \tilde{H}(R_c) &= H(R_c)\\
         \frac{\delta \tilde{\Psi}}{\tilde{\Psi}_0'}(R_c) &= \frac{\delta \Psi}{\Psi_0'}(R_c),
    \end{eqnarray}
    where, $R_c$ represents the radius of the boundary. Physical quantities with no tilde represent their value in the SF region just below the junction. A tilde represents the value of the physical quantity in the normal fluid region just above the junction.
    
\subsection{Odd parity sector}
\label{junction condition for odd}

For the continuity of the quantities of the odd mode perturbation we follow  similar procedure. But as has been discussed earlier we consider the time dependent perturbation for that purpose. we find 
    
    \begin{eqnarray}
         \gamma_{03} &= \delta g_{03}\\
         \gamma_{13} &= -\delta g_{13}\\
         K_{03} &= \frac{1}{\sqrt{g^{(0)11}}}(\dot{\delta g_{13}}-\delta g_{03}')
    \end{eqnarray}
    
    Taking $h_i (t,r) = \int d\omega \hat{h}_i (\omega,r)e^{-i\omega t}$ implies $\hat{h}_i$ is continuous implying $\psi$ is continuous (for the definition check \ref{odd perturbation}). Continuity of $K_{03}$ implies $\omega e^{(\nu-\kappa)/2}r\psi + \frac{\{e^{(\nu-\kappa)/2}(\psi r)'\}'}{\omega}$ is continuous. Using Eq.(\ref{omega psi equation}) in SF region we find that the following expression is continuous:

    \begin{equation}
    \begin{split}
        \omega e^{(\kappa-\nu)/2}r\psi + \frac{e^{(\nu-\kappa)/2}}{\omega}\big[2\psi' +\psi e^{\kappa} \big(-\frac{4M(r)}{r^2}+\frac{l(l+1)}{r}\big)\big].
    \end{split}
    \end{equation}

    A similar expression can be found in the normal fluid region with $\Psi \rightarrow p$ and $\Lambda \rightarrow -\rho$.
     Since $\psi$ is continuous, this implies $\psi'$ is continuous across the junction.

\section{Expressions for matter variables}\label{Expressions for matter variables}
\begin{widetext}

In the limit $K \rightarrow 0$ the master function and the chemical potentials of the neutron and proton fluids can be expressed as, 
\begin{eqnarray}
    \left.\Lambda\right|_0 &=&  - \frac{c_\omega^2}{18 \pi^4}\left(k_n^3+k_p^3\right)^2-\frac{c_\rho^2}{72 \pi^4}\left(k_p^3-k_n^3\right)^2- \frac{1}{4 \pi^2} \left(k_n^3 
             \sqrt{k_n^2 + \left.m^2_*\right|_0} + 
             k_p^3 \sqrt{k_p^2 + \left.m^2_*\right|_0}
             \right)   \cr
             && \cr
             &&- \frac{1}{4} c_\sigma^{- 2} \left[\left(2 m - \left.m_*\right|_0\right) 
             \left(m - \left.m_*\right|_0\right)+\left.m_*\right|_0\left( b m c_\sigma^2\left(m-\left.m_*\right|_0\right)^2 
          + c  c_\sigma^2\left(m-\left.m_*\right|_0\right)^3\right)\right]\nonumber \\
          &&-\frac{1}{3} b m \left(m-\left.m_*\right|_0\right)^3\--\frac{1}{4} c  \left(m-\left.m_*\right|_0\right)^4- \frac{1}{8 \pi^2} \left( k_p \left[2 k_p^2 + m_e^2\right] \sqrt{k_p^2 + m^2_e}\right.\nonumber \\
             &&
             \left. - m^4_e {\rm ln}\left[
             \frac{k_p + \sqrt{k_p^2 + m^2_e}}{m_e}\right]\right) 
              \ , \\
             && \cr
    \left.\mu\right|_0 &=&-\frac{\pi^2}{k_n^2} \left.\frac{\partial \Lambda}{\partial k_n}\right|_0= \frac{c_\omega^2}{3 \pi^2}\left(k_n^3+k_p^3\right)
 - \frac{c_\rho^2}{12 \pi^2}\left(k_p^3-k_n^3\right)  + \sqrt{k_n^2 + 
             \left.m^2_*\right|_0} \ , \\ 
             && \cr
    \left.\chi\right|_0 &=&-\frac{\pi^2}{k_p^2} \left.\frac{\partial \Lambda}{\partial k_p}\right|_0= \frac{c_\omega^2}{3 \pi^2}\left(k_n^3+k_p^3\right)
+ \frac{c_\rho^2}{12 \pi^2}\left(k_p^3-k_n^3\right) + \sqrt{k_p^2 + 
             \left.m^2_*\right|_0} + \sqrt{k_p^2 + m_e^2}. \  
\end{eqnarray}             
The generalized pressure $\Psi$ and the master function are realized by the following relationship,
\begin{equation}
    \left.\Psi\right|_0 = \left.\Lambda\right|_0 + \frac{1}{3 \pi^2}
             \left(\left.\mu\right|_0 k_{n}^3 + 
             \left.\chi\right|_0 k_{p}^3\right)~.
\end{equation} 
In the above expressions, $c_{\sigma}^2 = (g_{\sigma}/m_{\sigma})^2$, $c_{\omega}^2=(g_\omega/m_\omega)^2$,  $c_{\rho}^2=(g_\rho/m_\rho)^2$ and
\begin{eqnarray}
\label{m star eqn}
    \left.m_*\right|_0 &=& m_*(k_n,k_p,0) \cr
         && \cr
        &=& m - \left.m_*\right|_0 \frac{c_\sigma^2}{2 \pi^2}  
            \left(k_n \sqrt{k_n^2 + \left.m^2_*\right|_0} + k_p 
            \sqrt{k_p^2 + \left.m^2_*\right|_0} + \frac{1}{2} 
            \left.m_*^2\right|_0 {\rm ln} \left[\frac{- k_n + 
            \sqrt{k_n^2 + \left.m^2_*\right|_0}}{k_n + 
            \sqrt{k_n^2 + \left.m^2_*\right|_0}}\right] \right. \cr
         && \cr
         && + \frac{1}{2} \left.\left.m_*^2\right|_0 {\rm ln} \left[\frac{- 
            k_p + \sqrt{k_p^2 + \left.m^2_*\right|_0}}{ k_p + 
            \sqrt{k_p^2 + \left.m^2_*\right|_0}}\right]\right) \ + b  m  c_\sigma^2\left(m - m_*\right)^2 +   c  c_\sigma^2\left(m - m_*\right)^3~. 
            \label{mstar}
\end{eqnarray}

The expressions for the other matter coefficients (see \cite{Kheto2014,Kheto2015}) that are used as the inputs in field equations are as follows:
\begin{eqnarray}
{\cal A}|_0 &=& c_{\omega}^2-\frac{1}{4} c_{\rho}^2 + \frac{c^2_{\omega} }{ 5 
        \left.\mu^2\right|_0} \left(2 k_p^2 \frac{\sqrt{k_n^2 + 
        \left.m^2_*\right|_0}}{\sqrt{k_p^2 + 
        \left.m^2_*\right|_0}} + \frac{c^2_{\omega}}{3 \pi^2} 
        \left[\frac{k_n^2 k_p^3}{\sqrt{k_n^2 + 
        \left.m^2_*\right|_0}} + \frac{k_p^2 k_n^3}{\sqrt{k_p^2 + 
        \left.m^2_*\right|_0} }\right]\right)\cr
         && \cr
         &&
        +\frac{c^2_{\rho} }{ 20 
        \left.\mu^2\right|_0} \left(2 k_p^2 \frac{\sqrt{k_n^2 + 
        \left.m^2_*\right|_0} }{\sqrt{k_p^2 + 
        \left.m^2_*\right|_0}} + \frac{c^2_{\rho} }{ 12 \pi^2} 
        \left[\frac{k_n^2 k_p^3 }{ \sqrt{k_n^2 + 
        \left.m^2_*\right|_0}} + \frac{k_p^2 k_n^3 }{\sqrt{k_p^2 + 
        \left.m^2_*\right|_0} }\right]\right)\cr
         && \cr 
         &&
         -\frac{c^2_{\rho}c^2_{\omega} }{ 30\left.\mu^2\right|_0 \pi^2} 
        \left[\frac{k_n^2 k_p^3 }{ \sqrt{k_n^2 + 
        \left.m^2_*\right|_0}} - \frac{k_p^2 k_n^3 }{ \sqrt{k_p^2 + 
        \left.m^2_*\right|_0} }\right]  + \frac{3 \pi^2 k_p^2 
        }{ 5 \left.\mu^2\right|_0 k_n^3} \frac{k_n^2 + 
        \left.m^2_*\right|_0 }{ \sqrt{k_p^2 + 
        \left.m^2_*\right|_0}} \ , \\
        && \cr
{\cal B}|_0 &=& \frac{3 \pi^2 \left.\mu\right|_0 }{ k_n^3} - 
        c_{\omega}^2 \frac{k_p^3 }{ k_n^3}+ \frac{1}{4}c_{\rho}^2 \frac{k_p^3 }{ k_n^3} - \frac{c^2_{\omega} k_p^3 
        }{ 5 \left.\mu^2\right|_0 k_n^3} \left(2 k_p^2 
        \frac{\sqrt{k_n^2 + \left.m^2_*\right|_0} }{ \sqrt{k_p^2 + 
        \left.m^2_*\right|_0}} + \frac{c^2_{\omega} }{ 3 \pi^2} 
        \left[\frac{k_n^2 k_p^3}{ \sqrt{k_n^2 + 
        \left.m^2_*\right|_0}} + \frac{k_p^2 k_n^3 }{ \sqrt{k_p^2 + 
        \left.m^2_*\right|_0} }\right]\right)  \cr
        && \cr
        && - \frac{c^2_{\rho} k_p^3 
        }{ 20 \left.\mu^2\right|_0 k_n^3} \left(2 k_p^2 
        \frac{\sqrt{k_n^2 + \left.m^2_*\right|_0} }{ \sqrt{k_p^2 + 
        \left.m^2_*\right|_0}} + \frac{c^2_{\rho} }{ 12 \pi^2} 
        \left[\frac{k_n^2 k_p^3}{ \sqrt{k_n^2 + 
        \left.m^2_*\right|_0}} + \frac{k_p^2 k_n^3 }{\sqrt{k_p^2 + 
        \left.m^2_*\right|_0} }\right]\right)  \cr
        && \cr
        && + \frac{c^2_{\rho} c^2_{\omega}k^3_p }{ 30 \pi^2\left.\mu^2\right|_0 k_n^3} 
        \left[\frac{k_n^2 k_p^3 }{ \sqrt{k_n^2 + 
        \left.m^2_*\right|_0}} - \frac{k_p^2 k_n^3 }{ \sqrt{k_p^2 + 
        \left.m^2_*\right|_0} }\right]-\frac{3 \pi^2 k_p^5 }{ 5 \left.\mu^2\right|_0 k_n^6} 
        \frac{k_n^2 + \left.m^2_*\right|_0 }{ \sqrt{k_p^2 + 
        \left.m^2_*\right|_0}}\,\label{b00} \ , \\
        && \cr
{\cal C}|_0 &=& \frac{3 \pi^2 \left.\chi\right|_0 }{ k_p^3}+ \frac{1}{4}c_{\rho}^2 \frac{k_n^3 }{ k_p^3} - 
        c_{\omega}^2 \frac{k_n^3}{k_p^3} - \frac{c^2_{\omega} k_n^3 
        }{ 5 \left.\mu^2\right|_0 k_p^3} \left(2 k_p^2 
        \frac{\sqrt{k_n^2 + \left.m^2_*\right|_0} }{ \sqrt{k_p^2 + 
        \left.m^2_*\right|_0}} + \frac{c^2_{\omega} }{ 3 \pi^2} 
        \left[\frac{k_n^2 k_p^3 }{ \sqrt{k_n^2 + 
        \left.m^2_*\right|_0}} + \frac{k_p^2 k_n^3 }{ \sqrt{k_p^2 + 
        \left.m^2_*\right|_0} }\right]\right)  \cr
        && \cr
        && - \frac{c^2_{\rho} k_n^3 
        }{ 20 \left.\mu^2\right|_0 k_p^3} \left(2 k_p^2 
        \frac{\sqrt{k_n^2 + \left.m^2_*\right|_0} }{ \sqrt{k_p^2 + 
        \left.m^2_*\right|_0}} + \frac{c^2_{\rho} }{ 12 \pi^2} 
        \left[\frac{k_n^2 k_p^3}{ \sqrt{k_n^2 + 
        \left.m^2_*\right|_0}} + \frac{k_p^2 k_n^3 }{ \sqrt{k_p^2 + 
        \left.m^2_*\right|_0} }\right]\right)\cr
        && \cr
        && +\frac{c^2_{\rho} c^2_{\omega}k_n^3 }{ 30 \pi^2\left.\mu^2\right|_0 k_p^3} 
        \left[\frac{k_n^2 k_p^3 }{ \sqrt{k_n^2 + 
        \left.m^2_*\right|_0}} - \frac{k_p^2 k_n^3 }{ \sqrt{k_p^2 + 
        \left.m^2_*\right|_0} }\right]- \frac{3 \pi^2 }{ 5 \left.\mu^2\right|_0 k_p} 
        \frac{k_n^2 + \left.m^2_*\right|_0 }{ \sqrt{k_p^2 + 
        \left.m^2_*\right|_0}}~,
\label{coo} 
\end{eqnarray}

\begin{eqnarray}
{{\cal A}_0^0}|_0 &=-& \frac{\pi^4}{k_p^2k_n^2} \left.\frac{\partial^2 \Lambda}{\partial k_p\partial k_n}\right|_0
        = c_\omega^2 - \frac{c_\rho^2}{4}+ \frac{\pi^2 }{ k^2_p} \frac{ 
        \left.m_*\right|_0 \left.\frac{\partial m_* }{ \partial k_p}
        \right|_0 }{ \sqrt{k^2_n + \left.m^2_*\right|_0}}\ , \\
        && \cr
{{\cal B}_0^0}|_0 &=& \frac{\pi^4}{k_n^5} \left(\left.2\frac{\partial \Lambda}{\partial k_n}\right|_0-k_n\left.\frac{\partial^2 \Lambda}{\partial k_n^2}\right|_0\right) 
        = c_\omega^2 + \frac{c_\rho^2}{4} + \frac{\pi^2 }{ k^2_n} \frac{k_n + 
        \left.m_*\right|_0 \left.\frac{\partial m_* }{ \partial k_n}
        \right|_0 }{ \sqrt{k^2_n + \left.m^2_*\right|_0}} \ , \\
        && \cr
{{\cal C}_0^0}|_0 &=& \frac{\pi^4}{k_p^5} \left(\left.2\frac{\partial \Lambda}{\partial k_p}\right|_0-k_p\left.\frac{\partial^2 \Lambda}{\partial k_p^2}\right|_0\right)
        = c_\omega^2 +  \frac{c_\rho^2}{4} + \frac{\pi^2 }{ k^2_p} \frac{k_p + 
        \left.m_*\right|_0 \left.\frac{\partial m_* }{ \partial k_p}
        \right|_0 }{ \sqrt{k^2_p + \left.m^2_*\right|_0}} + 
        \frac{\pi^2 }{ k_p} \frac{1 }{ \sqrt{k^2_p + m^2_e}},
\end{eqnarray}
where,
\begin{eqnarray}
     \left.\frac{\partial m_* }{ \partial k_n}\right|_0 &=& - 
          \frac{c_\sigma^2 }{ \pi^2} \frac{\left.m_*\right|_0 k_n^2 
          }{ \sqrt{k_n^2 + \left.m^2_*\right|_0}} \left(\frac{3 m - 2 
          \left.m_*\right|_0 +3 b m c_\sigma^2\left(m-\left.m_*\right|_0\right)^2 
          +3 c  c_\sigma^2\left(m-\left.m_*\right|_0\right)^3}{\left.m_*\right|_0}\right.\nonumber \\ 
          &&\left.- \frac{c_\sigma^2 
          }{\pi^2} \left[\frac{k_n^3 }{ \sqrt{k_n^2 + \left.m^2_*
          \right|_0}} + \frac{k_p^3 }{ \sqrt{k_p^2 + 
          \left.m^2_*\right|_0}}\right]+2 b m c_\sigma^2\left(m-\left.m_*\right|_0\right) 
          +3 c  c_\sigma^2\left(m-\left.m_*\right|_0\right)^2\right)^{- 1}, 
          \\
 \textnormal{and,} \nonumber \\
 	&& \cr
     \left.\frac{\partial m_* }{ \partial k_p}\right|_0 &=& - 
          \frac{c_\sigma^2 }{ \pi^2} \frac{\left.m_*\right|_0 k_p^2 
          }{ \sqrt{k_p^2 + \left.m^2_*\right|_0}} \left(\frac{3 m - 2 
          \left.m_*\right|_0 +3 b m c_\sigma^2\left(m-\left.m_*\right|_0\right)^2 
          +3 c  c_\sigma^2\left(m-\left.m_*\right|_0\right)^3}{\left.m_*\right|_0}\right.\nonumber \\ 
          &&\left.- \frac{c_\sigma^2 
          }{\pi^2} \left[\frac{k_n^3 }{ \sqrt{k_n^2 + \left.m^2_*
          \right|_0}} + \frac{k_p^3 }{ \sqrt{k_p^2 + 
          \left.m^2_*\right|_0}}\right]+2 b m c_\sigma^2\left(m-\left.m_*\right|_0\right) 
          +3 c  c_\sigma^2\left(m-\left.m_*\right|_0\right)^2\right)^{- 1}, 
\end{eqnarray}
respectively.
\end{widetext}


\begin{thebibliography}{99}
\bibitem{Abbott2017} B. P. Abbott {\it et al.} (LIGO Scientific and Virgo Collaborations), Phys. Rev. Lett. {\bf 119}, 161101 (2017).
\bibitem{Abbott2018} B. P. Abbott {\it et al.} (LIGO Scientific and Virgo Collaborations), Phys. Rev. Lett. {\bf 121}, 161101 (2018).



\bibitem{Damour2009} T. Damour, A. Nagar, Phys. Rev. D {\bf80}, 084035 (2009)
\bibitem{Binnington2009} T. Binnington, E. Poisson Phys. Rev. D {\bf 80}, 084018 (2009)
\bibitem[Hinderer et al. (2010)]{Hinderer2010} T. Hinderer, B. D. Lackey, R. N. Lang and J. S. Read, Phys. Rev. D {\bf 81}, 123016 (2010).
\bibitem{Flanagan2008}\'{E}. \'{E} Flanagan, and T. Hinderer, Phys. Rev. D {\bf 77}, 021502 (2008).
\bibitem{Hinderer2008} T. Hinderer, Astrophys. J. {\bf 677}, 1216 (2008).

\bibitem{Agathos2015} M.  Agathos,  J.  Meidam,  W.  Del  Pozzo,  T.  G.  F.  Li, M. Tompitak, J. Veitch, S. Vitale,  and C. Van Den Broeck, Phys. Rev. D {\bf 92} , 023012 (2015).
\bibitem{Takami2014} K. Takami, L. Rezzolla, and L. Baiotti, Phys. Rev. Lett. {\bf 113}, 091104 (2014).

\bibitem{Bose2018}S. Bose, K. Chakravarti, L. Rezzolla, B. S. Sathyaprakash, K. Takami, Phys. Rev. Lett. {\bf 120}, 031102 (2018)


\bibitem{Vines2011} J. Vines, E. E. Flanagan, and T. Hinderer, Phys. Rev. D {\bf 83}, 084051 (2011).

\bibitem{Damour2012} T. Damour, A. Nagar, and L. Villain, Phys. Rev. D {\bf 85}, 123007 (2012) 

\bibitem{Read2013} J. S. Read, L. Baiotti, J. D. E. Creighton, J. L. Friedman, B. Giacomazzo, K. Kyutoku, C. Markakis, L. ezzolla, M. Shibata, and K. Taniguchi, Phys. Rev. D {\bf 88}, 044042 (2013).

\bibitem{DelPozzo2013}W. Del Pozzo, T. G. F. Li, M. Agathos, C. Van Den Broeck, and S. Vitale, Phys. Rev. Lett. {\bf 111}, 071101 (2013).

\bibitem{Wade2014}L. Wade, J. D. E. Creighton, E. Ochsner, B. D. Lackey, B. F. Farr, T. B. Littenberg, and V. Raymond, Phys. Rev. D {\bf 89} , 103012 (2014)

\bibitem{Favata2014} 
M. Favata, 
Phys. Rev. Lett. {\bf 112}, 101101 (2014).
 
 \bibitem{Hotokezaka2016} 
K. Hotokezaka, K. Kyutoku, Y. Sekiguchi, and M. Shibata, 
Phys. Rev. D {\bf 93}, 064082 (2016)


\bibitem{Sedrakian2018} A. Sedrakian and J. W. Clark, arXiv:1802.00017v1 [nucl:th]

\bibitem{Clark1992}J. Clark, R. Dav\'e, and J. Chen, The Structure and Evolution of Neutron Stars, edited by D. Pines, R. Tamagaki, and S. Tsurate (Addison-Wesley, Redwood City, CA, 1992).




\bibitem{Migdal1959}A. B. Migdal, Nucl. Phys. {\bf 13}, 655 (1959);


\bibitem{Baym1975} G. Baym, C. J. Pethick, D. Pines, and M. Ruderman, Nature (London) {\bf 224}, 872 (1969).

\bibitem{Anderson1975} P. W. Anderson and N. Itoh, Nature (London) {\bf 256}, 25 (1975)

\bibitem{Page2011} D. Page, M. Prakash, J. M. Lattimer, and A. W. Steiner, Phys. Rev. Lett. {\bf 106}, 081101 (2011)

\bibitem{Shternin2011} P. S. Shternin, D. G. Yakovlev, C. O. Heinke, W. C. G. Ho, and D. J. Patnaude, Mon. Not. R. Astron. Soc. {\bf 412}, L108 (2011);

\bibitem{Char2018}
Prasanta Char, Sayak Datta,
Phys. Rev. D 98, 084010 (2018).

\bibitem{Carter1989} B. Carter, Relativistic Fluid Dynamics, edited by A. Anile and M. Choquet-Bruhat ~Springer-Verlag, Berlin, 1989.
\bibitem{Comer1994} G.L. Comer and D. Langlois, Class. Quantum Grav. {\bf 11}, 709 (1994).
\bibitem{Carter1995}  B. Carter and D. Langlois, Phys. Rev. D {\bf 51}, 5855 (1995).
\bibitem{carter1998_1} B. Carter and D. Langlois, Nucl. Phys. {\bf B454}, 402 (1998).
\bibitem{carter1998_2} B. Carter and D. Langlois, Nucl. Phys. {\bf B531}, 478 (1998).
\bibitem{Langois1998} D. Langlois, A. Sedrakian, and B. Carter, Mon. Not. R. Astron. Soc. {\bf 297}, 1189 1998.
\bibitem{Prix2000} R. Prix, Phys. Rev. D {\bf 62}, 103005 (2000).


\bibitem{Andersson_2002}
N. Andersson, G. L. Comer, and D. Langlois
Phys. Rev. D 66, 104002 (2002).


\bibitem{Biswas2019elasticity}
B. Biswas, R. Nandi, P. Char and S. Bose, 
Phys. Rev. D 100, 044056 (2019).


\bibitem{Andersson2001} N. Andersson and G.L. Comer, Class. Quantum Grav. {\bf 18}, 969 (2001).
\bibitem{Comer1999} 
G. Comer, D. Langlois and L. M. Lin, 
Phys. Rev.  D {\bf 60}, 104025 (1999).

\bibitem{Comer2003}
G. L. Comer and R. Joynt, 
Phy. Rev. D {\bf 68}, 023002 (2003).

\bibitem{Comer2004}
G. Comer,
Phys. Rev. D {\bf 69}, 123009 (2004).

\bibitem{Kheto2014}
A. Kheto and D. Bandyopadhyay, 
Phys. Rev. D {\bf 89}, 023007 (2014).

\bibitem{Kheto2015}
A. Kheto and D. Bandyopadhyay,
Phys. Rev. D {\bf 91}, 043006 (2015).

\bibitem{grill2014}
F. Grill, H. Pais, C. Provid\^encia, I. Vida\~na, and S. S. Avancini, 
Phys. Rev. C {\bf 90}, 045803 (2014)

\bibitem{haensel2007}
Haensel, P., Potekhin, A. Y.,  Yakovlev, D. G. 2007, Neutron Stars 1,
Equation of State and Structure (Berlin: Springer)


\bibitem{Thorne1967} K. S. Thorne and A. Campolattaro, Astrophys. J. {\bf 149}, 591 (1967)


\bibitem{Chandra}
S. Chandrasekhar, 
\textit{The Mathematical Theory of Black Holes}~(Oxford University press, New Delhi 2010)


\bibitem{Regge1957}
T. Regge and J. A. Wheeler,
Phys. Rev. {\bf 108}, 1063 (1957).


\bibitem{Pani_2018}
Paolo Pani, Leonardo Gualtieri, Tiziano Abdelsalhin, and Xisco Jiménez-Forteza,
Phys. Rev. D 98, 124023 (2018).

\bibitem{Fattoyev2010} F.J. Fattoyev, C.J. Horowitz, J. Piekarewicz, and G. Shen, Phys. Rev. C {\bf 82}, 055803 (2010).

\bibitem{Glendenning1991}N. K. Glendenning and S. A. Moszkowski, Phys. Rev. Lett. {\bf 67}, 2414 (1991).


\bibitem{Yagi_2014}
Kent Yagi
Phys. Rev. D 89, 043011 (2014).



\bibitem{Maselli_2013}
Andrea Maselli, Vitor Cardoso, Valeria Ferrari, Leonardo Gualtieri, and Paolo Pani
Phys. Rev. D 88, 023007 (2013)


\bibitem{Nandi_2018}
Rana Nandi, Prasanta Char, Subrata Pal
arXiv:1809.07108

\bibitem{Bharat_2019}
Bharat Kumar, Philippe Landry
	arXiv:1902.04557
	
	
\bibitem{Zack_2019}
Zack Carson, Andrew W. Steiner, and Kent Yagi
Phys. Rev. D 99, 043010 (2019).



\bibitem{Zack_2019_2}
Zack Carson, Katerina Chatziioannou, Carl-Johan Haster, Kent Yagi, Nicolás Yunes,
	arXiv:1903.03909
	
	
\bibitem{Luca_2019}
Luca Baiotti, arXiv:1907.08534

\bibitem{Biswas2019anisotropy}
B. Biswas and S. Bose, arXiv:1903.04956 [gr-qc].

\bibitem{Raposo2019}
Guilherme Raposo, Paolo Pani, Miguel Bezares, Carlos Palenzuela, and Vitor Cardoso
Phys. Rev. D 99, 104072 (2019).

	
\bibitem{MTW}
C.W. Misner, K.S. Thorne and J.A. Wheeler, Gravitation (Freeman Press, San Fransisco, 1973)



	





\end{thebibliography}
\end{document}